\documentclass[acmsmall]{acmart}

\AtBeginDocument{%
  \providecommand\BibTeX{{%
    \normalfont B\kern-0.5em{\scshape i\kern-0.25em b}\kern-0.8em\TeX}}}

\setcopyright{acmcopyright}
\copyrightyear{2021}
\acmYear{2021}
\acmDOI{10.1145/3436892}

\acmJournal{TKDD}
\acmVolume{15}
\acmNumber{4}
\acmArticle{61}
\acmMonth{5}

\usepackage{graphics}
\usepackage{subfigure}
\usepackage{multirow}
\usepackage{amsmath}
\usepackage{bm}
\usepackage{algorithm}
\usepackage{algorithmic}
\usepackage{color}

\newcommand{\sgn}{\text{sgn}}
\begin{document}

\newcommand{\jy}[1]{\textcolor{black}{#1}}
\newcommand{\dk}[1]{\textcolor{black}{#1}}

\title{Search Efficient Binary Network Embedding}


\author{Daokun Zhang}
\affiliation{%
	\institution{Discipline of Business Analytics, The University of Sydney}
	\city{NSW 2006}
	\country{Australia}}
\email{Daokun.Zhang@sydney.edu.au}

\author{Jie Yin}
\affiliation{%
  \institution{Discipline of Business Analytics, The University of Sydney}
  \city{NSW 2006}
  \country{Australia}}
\email{jie.yin@sydney.edu.au}

\author{Xingquan Zhu}
\affiliation{%
 \institution{Department of Computer \& Electrical Engineering and Computer Science, Florida Atlantic University}
 \city{Boca Raton}
 \country{USA}}
\email{xzhu3@fau.edu}

\author{Chengqi Zhang}
\affiliation{%
  \institution{Australian Artificial Intelligence Institute, Faculty of Engineering and Information Technology, University of Technology Sydney}
  \city{Sydney}
  \country{Australia}}
\email{Chengqi.Zhang@uts.edu.au}

\renewcommand{\shortauthors}{Zhang, et al.}

\begin{abstract}
Traditional network embedding primarily focuses on learning a continuous vector representation for each node, preserving network structure and/or node content information, such that off-the-shelf machine learning algorithms can be easily applied to the vector-format node representations for network analysis. However, the learned continuous vector representations are inefficient for large-scale similarity search, which often involves finding nearest neighbors measured by distance or similarity in a continuous vector space. In this paper, we propose a search efficient binary network embedding algorithm called BinaryNE to learn a binary code for each node, by simultaneously modeling node context relations and node attribute relations through a three-layer neural network. BinaryNE learns binary node representations through a stochastic gradient descent based online learning algorithm. The learned binary encoding not only reduces memory usage to represent each node, but also allows fast bit-wise comparisons to support faster node similarity search than using Euclidean distance or other distance measures. Extensive experiments and comparisons demonstrate that BinaryNE not only delivers more than 25 times faster search speed, but also provides comparable or better search quality than traditional continuous vector based network embedding methods. \jy{The binary codes learned by BinaryNE also render competitive performance on node classification and node clustering tasks.} The source code of this paper is available at https://github.com/daokunzhang/BinaryNE.

\end{abstract}



\maketitle

\section{Introduction}\label{sec:introduction}

Networks offer a natural way to capture intrinsic relationships between entities -- social interactions among people, collaborations between co-workers, biological interactions among proteins, flow-of-funds between financial transactions, and so on. Networks can be modeled as a graph, where nodes indicate entities and edges indicate pairwise relationships between entities. Searching similar nodes in networks is an essential network analytic task, which directly benefits many real-world applications. For social security, potential terrorists can be detected by searching people with the same organization associations in the communication networks or online social networks. On e-commerce platforms, personalized recommendations can be effectively delivered by searching users with similar interests among users' social relations. In social networks, social actors with important structural roles, such as the center of a star or the hole spanner, can be discovered by searching nodes with the same properties among the whole network. Searching similar nodes can also benefits other tasks, such as Web page retrieval in the World Wide Web~\cite{haveliwala2002topic}, link prediction in social networks~\cite{srilatha2016similarity}, and identity resolution in bibliographic collaboration networks~\cite{zhang2015panther}.


To enable similarity search over networks, structural properties including common neighbors and structural context have been leveraged to estimate the similarity between nodes. Representative algorithms include Personalized PageRank~\cite{haveliwala2002topic}, SimRank~\cite{jeh2002simrank}, P-Rank~\cite{zhao2009p}, TopSim~\cite{lee2012top}, and Panther~\cite{zhang2015panther}. However, these methods suffer from the following major drawback:

\begin{itemize}
	\item\textbf{Incapable of capturing node content similarity}.  In addition to network structure, network nodes are often associated with rich content, such as user profiles in social networks, texts in Web page networks. Node content contains crucial information that provides direct evidence to measure node similarity. The structure based similarity search methods fail to leverage the similarity measured by node content, leading to suboptimal search results.
\end{itemize}

Recently, network embedding~\cite{perozzi2014deepwalk,tang2015line,yang2015network,zhang2017network} has been proposed to facilitate network analytic tasks, which aims to embed network nodes into a low-dimensional continuous vector space, by preserving network structure and/or node content information. After learning new node representations, network analytic tasks can be easily carried out by applying off-the-shelf machine learning algorithms to the new embedding space. However, such a machine learning driven network embedding paradigm often results in node representations that are inefficient for large-scale similarity search in terms of both time and memory. Consider a network with 10 million nodes, if we learn 200-dimensional continuous vector-format representations for each node, it requires 15G memory to accommodate these representations using standard double precision numbers, which is prohibitively intractable for general computing devices. Given a query node, if we want to find its similar nodes among the whole network using Euclidean distance in the continuous embedding space, it requires 2 billion times of floating-point product operation and 2 billion times of floating-point addition operation, to measure distance between the query node and all other nodes in the network. The high computational cost makes it unsuitable for real-time retrieval systems that require responsive solutions. In summary, searching similar nodes with continuous node representations inevitably incurs high time and memory cost, resulting in unsatisfactory performance on large-scale networks.

As an alternative way of nearest neighbor search, 
locality-sensitive hashing techniques~\cite{chi2017hashing,rajaraman2011mining,zhao2014locality,rachkovskij2017binary} have been proposed to improve search efficiency. They transform the numeric vector data into binary (or integer) codes that preserve the similarity in the original space. As a consequence, data can be stored with low memory cost and similarity search can be conducted efficiently by calculating the Hamming distance, e.g., between node binary codes with bit-wise operations. Borrowing the idea of hashing, we propose to learn binary representations for network nodes, i.e., transforming network nodes into binary codes rather than numeric vectors, such that the memory and time efficiency for similarity search can be significantly improved. Despite its potential, the binary node representation learning is confronted with the following two challenges:
\begin{itemize}
	\item \textbf{Heterogeneity}. To guarantee search accuracy, binary node representations are expected to capture the information from both network structure and node content, i.e., preserving node similarity at both structure level and node content level. However, network structure and node content are not always consistent or correlated with each other. How to fuse information from these two heterogeneous sources into binary codes and make them complement rather than deteriorate each other is a big challenge. 
	\item \textbf{Scalability}. To learn binary node representations that well preserve network structure and node content, we need to jointly optimize the respective objectives, such as random walk context node prediction~\cite{perozzi2014deepwalk} and node attribute reconstruction~\cite{yang2017properties}, which are usually non-linear functions of node representations.  With the constraint that each dimension should take value from $\{0, 1\}$, the problem of finding the optimal solution to node binary representations that optimize the nonlinear objective function is a nonlinear integer programming problem, which has been proved NP-hard~\cite{hemmecke2010nonlinear}. When it comes to large-scale networks with millions or billions of nodes/edges, and high dimensional node content features, it is impossible to find the exact optimal solutions in an efficient way. To make the learning highly scalable, in the promise of assuring the quality of solutions, approximation techniques together with online or parallel learning strategies need to be developed.
\end{itemize}

An intuitive solution to binary network embedding is to first learn continuous node representations and then binarize them into binary codes with the conventional hashing techniques. However, because converting continuous embeddings into binary codes inevitably causes information loss, the learned binary codes cannot accurately capture node similarity at both structure and content level. As a result, as demonstrated later in our experiments, this two-step learning strategy usually results in suboptimal search accuracy. 

\dk{The binary network embedding algorithm can also be addressed by the supervised hash learning algorithms~\cite{zhu2016deep, cao2017hashnet} originally designed for non-relational data, with the required node similarity labels estimated from network structure and/or node attributes in advance. However, directly applying the supervised hash learning algorithms~\cite{zhu2016deep, cao2017hashnet} to the large-scale networks is confronted with three key challenges: 1) the high complexity for calculating all pairwise node similarity values from network structure with the objective of preserving high-order proximities as well as node attributes that are usually high-dimensional; 2) the information loss caused by quantizing the continuous similarity values into binary similarity labels; 3) most seriously, the imbalanced similarity relations, i.e., the number of dissimilar node pairs is overwhelmingly
larger than the number of similar node pairs on graphs.}

In this paper, we propose a novel Binary Network Embedding algorithm, called BinaryNE, to learn binary node representations directly from both network structure and node content features for efficient similarity search. BinaryNE learns binary node representations by simultaneously modeling node context and node attribute relations through a three-layer neural network, with the objective of capturing node similarity in both network structure and node content. To obtain binary codes, the \textit{sign} function $\sgn(\cdot)$ is employed as the activation function in the hidden layer. However, as the gradient of the \textit{sign} function is zero almost everywhere, traditional gradient decent based optimization strategies are infeasible for learning parameters, which is known as the \textit{ill-posed gradient} problem. To address this problem, we adopt the state-of-the-art continuation technique~\cite{allgower2012numerical, cao2017hashnet} and develop an online stochastic gradient descent algorithm to learn parameters, which guarantees the great scalability of BinaryNE. \dk{In the learning process, node binary representations are sequentially updated with regards to the given node-context or node-attribute occurrence pair, by maximizing the occurrence probability of the context node or node attribute conditioned on the given node. Reflected on the binary representation space, nodes are dragged by their context nodes and attributes back-and-forth until an equilibrium status. In this way, nodes sharing overlapping sets of context nodes and attributes are finally represented closely in the binary embedding space. Our learning strategy can effectively capture the similarity in network structure and node attributes, and avoid the learning process to be dominated by the overwhelmingly large quantities of dissimilar node pairs to the most extent, which seriously challenges the traditional supervised hash learning algorithms~\cite{zhu2016deep, cao2017hashnet}.} Experiments on six real-world networks show that BinaryNE exhibits much lower memory usage and quicker search speed than the state-of-the-art network embedding methods, while achieving appealing search accuracy. \jy{BinaryNE's binary node representations also deliver competitive results on node classification and node clustering tasks.}

The main contribution of this paper is threefold:
\begin{itemize}
	\item We analyze the feasibility and advantage of learning binary node representations as a solution to efficient node similarity search over large-scale networks involving node attributes.
	\item We propose a new algorithm called BinaryNE to effectively learn high-quality binary node representations from both network structure and node features, together with an efficient stochastic gradient descent based solution.
	\item \jy{Extensive experiments on six real-world networks validate the superiority of BinaryNE on similarity search in terms of search accuracy, memory usage and efficiency, as well as its competitive performance on node classification and node clustering.}
\end{itemize}

The remainder of this paper is organized as follows. In Section 2, we review the related work, including network embedding and node similarity search. In Section 3, we give a formal definition of binary network embedding and review the DeepWalk algorithm as preliminaries. The proposed BinaryNE algorithm is described in Section 4, followed by experiments presented in Section 5. Finally, we conclude this paper in Section 6.

\section{Related Work}
In this section, we review two lines of related work: network embedding that aims to learn node vector-format representations, and node similarity search that is realized by directly estimating node similarity from network structure. 

\subsection{Network Embedding}
According to whether the learned node representations take continuous or discrete values, the network embedding techniques can be divided into two groups: continuous network embedding and discrete network embedding.

\subsubsection{Continuous Network Embedding}
Depending on whether node content features are leveraged, continuous network embedding techniques can be divided into two groups: \textit{structure preserving network embedding} and \textit{attributed network embedding}. 

Structure preserving network embedding learns node representations from only network structure. DeepWalk~\cite{perozzi2014deepwalk} first encodes network structure into a set of random walk sequences, and then employs Skip-Gram~\cite{mikolov2013efficient} to learn node representations that capture structural context similarity. node2vec~\cite{grover2016node2vec} extends DeepWalk to better balance the local structure preserving and global structure preserving objective by leveraging biased random walks. LINE~\cite{tang2015line} learns node representations through directly modeling the first-order proximity (the proximity between connect nodes) and the second-order proximity (the proximity between nodes sharing direct neighbors). \dk{Although DeepWalk, node2vec and LINE are all based on the Skip-Gram model explicitly or implicitly, \cite{qiu2018network} proves their equivalence to a unified matrix factorization formulation.} GraRep~\cite{cao2015grarep} further extends LINE~\cite{tang2015line} to capture high-order proximities through the matrix factorization version of Skip-Gram~\cite{levy2014neural}. M-NMF~\cite{wang2017community} complements the local structure proximity with the intra-community proximity to learn community-aware node representations. DNGR~\cite{cao2016deep} first obtains high-dimensional structure preserving node representations through the proposed random surfing method, and then utilizes the \textit{stacked denoising autoencoder} (SDAE)~\cite{vincent2010stacked} to learn low-dimensional representations. SNDE~\cite{wang2016structural} employs deep autoencoder to learn deep nonlinear node representations, by reconstructing node adjacent matrix representations to preserve the second-order proximity and penalizing the representation difference of connected nodes to preserve the first-order proximity.

Attributed network embedding learns node representations by coupling node attributes with network structure. TADW~\cite{yang2015network} first proves the equivalence between DeepWalk~\cite{perozzi2014deepwalk} and a matrix factorization formulation, and then proposes to incorporate rich node text features into network embedding through inductive matrix factorization~\cite{natarajan2014inductive}. Through penalizing the distance of connected nodes in the embedding space, HSCA~\cite{zhang2016homophily} enforces TADW with the first-order proximity to obtain more informative node representations. UPP-SNE~\cite{zhang2017user} learns node representations by performing a structure-aware non-linear mapping on node content features. CANE~\cite{tu2017cane} learns context-aware node embeddings by applying the mutual attention mechanism on the attributes of connected nodes. MVC-DNE~\cite{yang2017properties} applies deep multi-view learning technique to fuse information from network structure and node content into node representations. GraphSAGE~\cite{hamilton2017inductive} first takes node content features as node representations, and then iteratively updates node representations by aggregating representations of neighboring nodes. AANE~\cite{huang2017accelerated} employs symmetric matrix factorization~\cite{kuang2012symmetric} to obtain node representations that capture attribute affinity, and simultaneously penalizes the representation difference between connected nodes. SINE~\cite{zhang2018sine} learns node representations for large-scale incomplete attributed networks by using node representations to simultaneously predict context nodes and node attributes. These network embedding algorithms learn task-general node representations in an unsupervised setting, where node class labels are not provided. 

Recently, several supervised network embedding algorithms have also been proposed, such as DMF~\cite{zhang2016collective}, TriDNR~\cite{pan2016tri}, DDRW~\cite{li2016discriminative}, MMDW~\cite{tu2016max}, LANE~\cite{huang2017label}, with the objective of learning discriminative node representations by exerting the power of available node labels. \jy{Built upon the success of deep neural networks on grid-structured data (e.g., images), graph neural networks (GNN), such as graph convolutional networks (GCN)~\cite{kipf2016semi} and graph attention networks (GAT)~\cite{velivckovic2017graph}, generalize to learning node embeddings on graphs by passing, transforming, and aggregating node features from the local neighborhood. The generated node embeddings can then be used as input to any differentiable prediction layer, e.g., for node classification or link prediction.}


\jy{All of the above network embedding techniques primarily consider embedding network nodes into a continuous Euclidean space, which could be favored by node classification tasks to achieve excellent performance. However, calculating pairwise similarity between nodes in a continuous space is computationally expensive, making it infeasible to perform node similarity search on large-scale networks.}

\subsubsection{Discrete Network Embedding} 

\dk{Graph based hashing~\cite{weiss2009spectral,liu2014discrete} has been proposed to learn binary codes for traditional non-relational data, such as images. This kind of hashing techniques work on the graphs constructed from data points by calculating their pairwise similarity. The constructed graph reflects the geographical data distribution in their original space, and helps learn similar binary codes for \jy{closely} located data points. The graphs considered by graph based hashing are artificially constructed from the input data, which are essentially different from real-world graphs \jy{where edges are naturally formed resulting from interactions between entities (nodes). In real-world natural graphs, connected nodes unnecessarily exhibit similar features~\cite{subbarai2015what,Bianconi2009Assessing}. This breaks the assumption made by graph based hashing~\cite{weiss2009spectral,liu2014discrete} that connected nodes are located closely in the Euclidean feature space. Therefore, directly applying graph based hashing to natural graphs may result in unsatisfactory performance. }}

Very recently, several embedding algorithms have been proposed to learn discrete node representations for real-world networks. For efficient node retrieval, Bernoulli Network Embedding~\cite{Misra2018bernoulli} learns binary node representations by modeling the generation of each dimension as a Bernoulli random test. \dk{KD-coding~\cite{chen2018learning} is proposed to learn discrete word embeddings with the Skip-Gram~\cite{mikolov2013efficient} model, which can also be coupled with DeepWalk~\cite{perozzi2014deepwalk} to learn discrete node representations. Through explicitly capturing high-order structure proximity, INH-MF~\cite{lian2018high} learns binary codes for network nodes with matrix factorization.} DNE~\cite{xiaobo2018discrete} learns binary node representations to speed up node classification. However, the above methods cannot support accurate search, because node content features are simply ignored. In addition, DNE is a supervised binary network embedding algorithm that requires node labels to be provided, which is different from our research that aims to learn binary node representations in an unsupervised setting. 

NetHash~\cite{wu2018efficient} is the first algorithm proposed to generate discrete node representations that encode both network structure and node content features. It applies the MinHash technique~\cite{broder2000min} to the union of tree-structured neighboring node features. As the learned discrete embeddings do not take binary values, similarity search with such embeddings tends to be inefficient. In this work, we aim to learn binary node representations that are directly optimized with binarization to enable similarity search efficacy and efficiency. BANE~\cite{yang2018binarized} is another algorithm that learns binary node representations from network structure and node attributes. BANE first constructs the Weisfeiler-Lehman proximity matrix~\cite{shervashidze2011weisfeiler} which enables neighboring node attribute vector aggregation and then performs binary factorization on the proximity matrix to obtain binary codes for each node. The Weisfeiler-Lehman matrix representation heavily relies the homophily assumption and tends to be less informative when neighboring nodes have discrepant attributes, which is, however, very common in real-world networks. Hence, BANE is less effective in retrieving similar nodes, as demonstrated later in the experiments. \dk{DGCN-BinCF~\cite{wang2019binarized} is proposed to learn binary node representations for bipartite networks formed by users and items together with their implicit feedback, with the purpose of binary collaborative filtering. To capture user-item relations implied by implicit feedback, GCN~\cite{kipf2016semi} is employed to learn their network embeddings. The structure information carried by network embeddings of users and items are then distilled into their binary codes to minimize the rank loss for recommendation. DGCN-BinCF is performed on the heterogeneous bipartite networks under the supervised recommendation setting, which is different from our problem of unsupervised binary node representation learning on homogeneous attributed networks. }

\subsection{Node Similarity Search}
To enable similarity search over networks, various metrics have been proposed to measure the structural relatedness between nodes. Bibliographic Coupling~\cite{kessler1963bibliographic} and Co-citation~\cite{small1973co} measure node similarity by counting the number of common neighbors. Other common neighbor based metrics include Jaccard's coefficient, Salton's coefficient, the Adamic/Acar coefficient~\cite{adamic2003friends}, \textit{etc}. This type of metrics are incapable of capturing the similarity between nodes sharing no common neighbors. SimRank~\cite{jeh2002simrank} estimates node similarity recursively with the principle that two nodes are similar if they have connections with similar nodes. Because calculating SimRank similarity is computationally expensive, other algorithms, like TopSim~\cite{lee2012top} and \cite{kusumoto2014scalable}, are proposed to reduce its time complexity. P-Rank~\cite{zhao2009p} enhances SimRank by jointly modeling both in- and out-link relationships for node structural similarity estimation. VertexSim~\cite{tsourakakis2014toward} represents each node as a convex combination of anchor nodes by optimizing a geometric objective, and then measures node similarity with the new representations. The above metrics only capture the similarity relying on the connectivity among the local neighborhood, but neglect the \textit{structural equivalence} between nodes sharing similar structural roles while being distantly located. \cite{jin2011axiomatic} justifies a series of axiomatic properties that should be satisfied by a role similarity measure, and proposes RoleSim, a role similarity measure, which is calculated in an iterative way and is proved to satisfy all the justified properties. Panther~\cite{zhang2015panther} estimates local structural similarity between pairwise nodes through their co-occurrence frequencies in randomly sampled paths. Panther++~\cite{zhang2015panther} augments Panther with structural role similarity by measuring the difference in neighboring node co-occurrence frequency distributions.

Calculating the aforementioned structure similarity metrics between all pairwise nodes, which is necessary for exact node similarity search, is time-consuming, with a time complexity at least quadratic to the number of nodes. Moreover, the above structure similarity metrics fail to capture the similarity measured by node features. The two limitations make the existing structural similarity estimation based search methods unsuitable for large-scale networks with rich node features.

\section{Problem Definition and Preliminaries}
In this section, we give a formal definition of the binary network embedding problem, followed by a review on the preliminaries of DeepWalk. 

\subsection{Problem Definition}
Assume we are given a network $G=(\mathcal{V},\mathcal{E},\mathcal{A},X)$, where $\mathcal{V}$ is the set of nodes, $\mathcal{E}\subseteq \mathcal{V}\times\mathcal{V}$ is the set of edges, and $\mathcal{A}$ is the set of attributes. $X\in\mathbb{R}^{|\mathcal{V}|\times|\mathcal{A}|}$ is the node feature matrix, with each element $X_{ij}\geq 0$ indicating the occurrence times/weights of attribute $a_{j}\in\mathcal{A}$ at node $v_{i}\in\mathcal{V}$. Here, we assume node attributes take the discrete values, which is natural for the text features in Web page/citation networks and most user profile features in social networks, like gender, affiliation, education type. In case of the numeric node attributes like age, we can easily discretize them into discrete/categorical values through transforming them into interval or bin based features.

The BinaryNE algorithm aims to learn binary representations for network nodes, i.e., learning a mapping function $\mathrm{\Phi}:v_{i}\in\mathcal{V}\mapsto\{+1,-1\}^{d}$, where $d$ is the dimension of the embedding space. The learned binary node representations $\Phi(v_{i})$ are expected to satisfy the following two properties: (1) \textbf{low-dimensional}: The dimension $d$ should be much smaller than the dimension of adjacent-matrix node representations, $|\mathcal{V}|$, the original high-dimensional node representations, for the sake of search efficiency; (2) \textbf{informative}: The learned binary node representations should capture node similarity measured by both network structure and node content features to guarantee the quality of node similarity search, 

\subsection{Preliminaries: DeepWalk}
Borrowing the idea of Skip-Gram model~\cite{mikolov2013efficient}, which learns word representations by preserving context similarity, DeepWalk leverages random walks to generate node context and represents nodes sharing similar context closely in the new embedding space. Given a random walk with length $L$, $\{v_{r_{1}}, v_{r_{2}},\cdots,v_{r_{i}},\cdots v_{r_{L}}\}$, for each node $v_{r_{i}}$, DeepWalk learns its representation by using it to predict its context nodes, which is realized by maximizing the occurrence probability of context nodes conditioned on this node:
\begin{equation}
\min_{\mathrm{\Phi}}-\log\mathrm{P}(\{v_{r_{i-t}}, \cdots, v_{r_{i+t}}\}\setminus v_{r_{i}}|v_{r_{i}}),
\end{equation}where $\{v_{r_{i-t}}, \cdots, v_{r_{i+t}}\}\setminus v_{r_{i}}$ are the context nodes of $v_{r_{i}}$ within a window size of $t$.  Here, the window size refers to the number of context nodes to be collected, which are located before or after the given target node in the given random walk.

Using the conditional independence assumption, the probability $\mathrm{P}(\{v_{r_{i-t}}, \cdots, v_{r_{i+t}}\}\setminus v_{r_{i}}|v_{r_{i}})$ can be calculated as
\begin{equation}
\mathrm{P}(\{v_{r_{i-t}}, \cdots, v_{r_{i+t}}\}\setminus v_{r_{i}}|v_{r_{i}}) = \prod_{j=i-t,j\neq i}^{i+t}\mathrm{P}(v_{r_{j}}|v_{r_{i}}).
\end{equation}Following~\cite{zhang2017user}, after a set of random walks are generated, we can formulate the overall optimization problem as
\begin{equation}
\min_{\mathrm{\Phi}} -\sum_{i=1}^{|\mathcal{V}|}\sum_{j=1}^{|\mathcal{V}|}n(v_{i},v_{j})\log\mathrm{P}(v_{j}|v_{i}),
\end{equation}where $n(v_{i},v_{j})$ is the occurrence time of node context pair $(v_{i},v_{j})$ collected from all random walks with $t$ window size and $\mathrm{P}(v_{j}|v_{i})$ is modeled by softmax:
\begin{equation}
\mathrm{P}(v_{j}|v_{i})=\frac{\exp(\mathrm{\Phi}(v_{i})\cdot \mathrm{\Psi}(v_{j}))}{\sum_{k=1}^{|\mathcal{V}|}\exp(\mathrm{\Phi}(v_{i})\cdot \mathrm{\Psi}(v_{k}))},\nonumber
\end{equation}where $\mathrm{\Psi}(\cdot)$ is the node embedding vector when the node act as a context node.

The overall optimization problem can be solved by iteratively sampling a node context pair $(v_{i},v_{j})$ and minimizing the following partial objective:
\begin{equation} \label{deepwalk_partial_obj}
\mathcal{O}_{ij}^{s}=-\log\mathrm{P}(v_{j}|v_{i}).
\end{equation}

\section{Binary Network Embedding}
This section details the optimization problem that we formulate for the binary network embedding, followed by the solution on how to solve it efficiently.

\begin{figure*}[t]
	\centering
	\includegraphics[width=0.8\columnwidth]{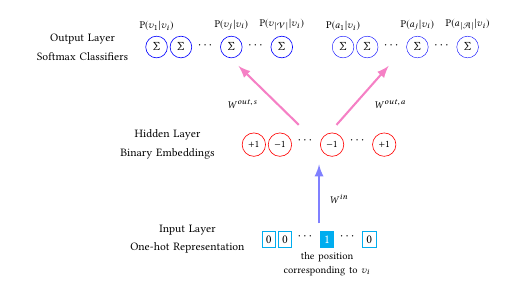} 
	\caption{The model architecture of BinaryNE. For each node $v_{i}$, BinaryNE learns its binary representation by using it to predict its context node $v_{j}$ and its attribute $a_{j}$.}
	\label{fig1:mechanism}
\end{figure*}

\subsection{The Optimization Problem}
Our objective is to learn informative binary network embeddings, with both network structure and node content features well preserved. The idea of DeepWalk can be borrowed for capturing network structure. To capture node content level similarity, we try to represent nodes sharing similar attributes closely in the low-dimensional space. \jy{To achieve this goal, we borrow the idea of Paragraph Vector~\cite{le2014distributed}, which learns document representations through modeling document-word relations with the Skip-Gram model. In this way, similar representations can be learned for documents with similar words. Here, to encode node attributes into binary node representations,}  we apply the idea of Skip-Gram~\cite{mikolov2013efficient} again, by using each node to predict its attributes. For each node attribute co-occurrence pair $(v_{i},a_{j})$, we minimize the following objective:
\begin{equation} \label{attribute_partial_obj}
\mathcal{O}_{ij}^{a}=-\log\mathrm{P}(a_{j}|v_{i}).
\end{equation}We illustrate the architecture of the proposed BinaryNE algorithm in Fig.~\ref{fig1:mechanism}, which is a three-layer neural network: the first layer is the one-hot representation for each node $v_{i}$, the hidden layer is the binary node representation $\mathrm{\Phi}(v_{i})\in\{+1,-1\}^{d}$ constructed from the input layer, and the output layer is the softmax conditional probability $\mathrm{P}(v_{j}|v_{i})$ and $\mathrm{P}(a_{j}|v_{i})$ for each context node $v_{j}$ and each attribute $a_{j}$, modeled through node binary representations in the hidden layer.

Given node $v_{i}$'s one-hot representation $\bm{p}^{i}\in\mathbb{R}^{|\mathcal{V}|}$  with 
$\bm{p}^{i}_{k}=1$ for $k=i$, and $\bm{p}^{i}_{k}=0$ for $k\neq i$. The binary node representation $\mathrm{\Phi}(v_{i})$ in the hidden layer is constructed by performing a linear transformation on $\bm{p}^{i}$ and activating it with the \textit{sign} function:
\begin{equation}\label{hidden}
\begin{aligned}
\mathrm{\Phi}(v_{i})&=\left[\sgn(\bm{p}^{i}\cdot W^{in}_{:1}),\sgn(\bm{p}^{i}\cdot W^{in}_{:2}),\cdots,\sgn(\bm{p}^{i}\cdot W^{in}_{:d})\right]^{\mathrm{T}}\\
&=\left[\sgn(W^{in}_{i1}),\sgn(W^{in}_{i2}),\cdots,\sgn( W^{in}_{id})\right]^{\mathrm{T}},
\end{aligned}
\end{equation}where $W_{:k}^{in}$ is the $k$-th column of $W^{in}\in\mathbb{R}^{|\mathcal{V}|\times d}$ (the weight matrix from the input layer to the hidden layer) and $\sgn(\cdot)$ is the \textit{sign} function, which is defined as
\begin{equation}\label{sgn_def}
\sgn(x) = \left\lbrace
\begin{aligned}
&+1,\;\mathrm{if}\;x\geqslant 0,\\
&-1,\;\mathrm{otherwise}.\nonumber
\end{aligned}
\right.
\end{equation}In the output layer, for the node context pair $(v_{i},v_{j})$, we model the probability $\mathrm{P}(v_{j}|v_{i})$ with softmax:
\begin{equation}\label{prob_context}
\mathrm{P}(v_{j}|v_{i})=\frac{\exp(\mathrm{\Phi}(v_{i})\cdot W^{out,s}_{:j})}{\sum_{k=1}^{|\mathcal{V}|}\exp(\mathrm{\Phi}(v_{i})\cdot W^{out,s}_{:k})},\nonumber
\end{equation}where $W^{out,s}_{:j}$ is the $j$-th column of $W^{out,s}\in\mathbb{R}^{d\times|\mathcal{V}|}$ (the weight matrix from the hidden layer to the output layer for predicting context node). Similarly, for the node attribute co-occurrence pair $(v_{i},a_{j})$, we model the probability $\mathrm{P}(a_{j}|v_{i})$ as
\begin{equation}\label{prob_content}
\mathrm{P}(a_{j}|v_{i})=\frac{\exp(\mathrm{\Phi}(v_{i})\cdot W^{out,a}_{:j})}{\sum_{k=1}^{|\mathcal{A}|}\exp(\mathrm{\Phi}(v_{i})\cdot W^{out,a}_{:k})},\nonumber
\end{equation}where $W^{out,a}_{:j}$ is the $j$-th column of $W^{out,a}\in\mathbb{R}^{d\times|\mathcal{A}|}$ (the weight matrix from the hidden layer to the output layer for predicting node attribute). 

To learn informative binary node embeddings, we integrate the structure proximity preserving objective in Eq. (\ref{deepwalk_partial_obj}) with the node attribute similarity preserving objective in Eq. (\ref{attribute_partial_obj}), and obtain the following overall optimization problem:
\begin{equation}\label{overall_obj}
\min_{\mathrm{\Phi}}\mathcal{O},
\end{equation}where
\begin{equation}\label{obj_function}
\mathcal{O}=-\alpha_{1}\sum_{i=1}^{|\mathcal{V}|}\sum_{j=1}^{|\mathcal{V}|}n(v_{i},v_{j})\mathrm{\log}\mathrm{P}(v_{j}|v_{i})
-\alpha_{2}\sum_{i=1}^{|\mathcal{V}|}\sum_{j=1}^{|\mathcal{A}|}X_{ij}\mathrm{\log}\mathrm{P}(a_{j}|v_{i}).
\end{equation}Here, $\alpha_{1}$ and $\alpha_{2}$ are the trade-off parameters to balance the contribution of the structure preserving objective and the node content preserving objective. Specifically, we set $\alpha_{1}$ and $\alpha_{2}$ to the reciprocal of the number of node context pairs and the number of observed node attribute co-occurrence pairs, respectively:
\begin{equation}
\alpha_{1}=\frac{1}{\sum_{i=1}^{|\mathcal{V}|}\sum_{j=1}^{|\mathcal{V}|}n(v_{i},v_{j})},\ \
\alpha_{2}=\frac{1}{\sum_{i=1}^{|\mathcal{V}|}\sum_{j=1}^{|\mathcal{A}|}X_{ij}},\nonumber
\end{equation} which essentially makes the objective Eq. (\ref{overall_obj}) perform minimization over the averaged $-\mathrm{\log}\mathrm{P}(v_{j}|v_{i})$ and $-\mathrm{\log}\mathrm{P}(a_{j}|v_{i})$. In Eq. (\ref{obj_function}), only the non-zero entries of $n(v_{i},v_{j})$ and $X_{ij}$ are considered, whose numbers are much smaller than $|\mathcal{V}|\times|\mathcal{V}|$ and $|\mathcal{V}|\times|\mathcal{A}|$, respectively.

\subsection{Solving the Optimization Problem}

As the derivative of the \textit{sign} function used to construct binary codes is zero almost everywhere, solving the optimization problem (\ref{overall_obj}) with gradient descent is \textit{ill-posed}. Following \cite{cao2017hashnet}, we approximate the non-smooth \textit{sign} function $\sgn(x)$ with its smooth proxy $\tanh(\beta x)$, which satisfies the following property:
\begin{equation}
\lim_{\beta\rightarrow\infty}\tanh(\beta x)=
\sgn(x).\nonumber
\end{equation}
\jy{Using $\tanh$}, node representation $\mathrm{\Phi}(v_{i})$ in Eq. (\ref{hidden}) is constructed as
\begin{equation}\label{hidden_new}
\mathrm{\Phi}(v_{i})=\left[\tanh(\beta W^{in}_{i1}),\tanh(\beta W^{in}_{i2}),\cdots,\tanh(\beta W^{in}_{id})\right]^{\mathrm{T}}.
\end{equation}With this continuous approximation, we can solve the optimization problem (\ref{overall_obj}) with stochastic gradient descent. At each iteration, we randomly select a node context pair $(v_{i},v_{j})$ according to the distribution of $n(v_{i},v_{j})$ or a node attribute co-occurrence pair $(v_{i},a_{j})$ according to the distribution of $X_{ij}$, and then update parameters towards minimizing the corresponding partial objective $\mathcal{O}_{ij}^{s}$ in Eq. (\ref{deepwalk_partial_obj}) or $\mathcal{O}_{ij}^{a}$ in Eq. (\ref{attribute_partial_obj}).

Given a sampled node context pair $(v_{i},v_{j})$, for training efficiency, we adopt negative sampling~\cite{gutmann2012noise} to approximate the partial objective $\mathcal{O}^{s}_{ij}$ in Eq. (\ref{deepwalk_partial_obj}) as
\begin{equation}
\mathcal{O}^{s}_{ij} = -\log\sigma(\mathrm{\Phi}(v_{i})\cdot W_{:j}^{out,s})
-\sum_{k:v_{k}\in\mathcal{V}_{neg}}\log\sigma(-\mathrm{\Phi}(v_{i})\cdot W_{:k}^{out,s}),
\end{equation}where $\mathcal{V}_{neg}$ is the set of sampled negative nodes and $\sigma(\cdot)$ is the sigmoid function. Then, we update the parameters with gradient descent:
\begin{equation}\label{update_structure}
\begin{aligned}
&W^{in}_{i:} = W^{in}_{i:}  - \eta \frac{\partial\mathcal{O}^{s}_{ij}}{\partial W^{in}_{i:}},\\
&W^{out,s}_{:j} = W^{out,s}_{:j}  - \eta \frac{\partial\mathcal{O}^{s}_{ij}}{\partial W^{out,s}_{:j}},\\
&W^{out,s}_{:k} = W^{out,s}_{:k}  - \eta \frac{\partial\mathcal{O}^{s}_{ij}}{\partial W^{out,s}_{:k}},\; \mathrm{for} \; v_{k}\in\mathcal{V}_{neg},\\
\end{aligned}
\end{equation}where $\eta$ is the learning rate. The gradients are calculated as
\begin{equation}
\begin{aligned}
&\frac{\partial\mathcal{O}_{ij}^{s}}{\partial W^{in}_{ir}} =
\beta [1-{\tanh(\beta W^{in}_{ir})}^{2}][\sigma(\mathrm{\Phi}(v_{i})\cdot W_{:j}^{out,s})-1]W_{rj}^{out,s}\\
&+\beta [1-{\tanh(\beta W^{in}_{ir})}^{2}] \sum_{k:v_{k}\in\mathcal{V}_{neg}}\sigma(\mathrm{\Phi}(v_{i})\cdot W_{:k}^{out,s})W_{rk}^{out,s},\\
&\frac{\partial\mathcal{O}_{ij}^{s}}{W^{out,s}_{:j}} = [\sigma(\mathrm{\Phi}(v_{i})\cdot W^{out,s}_{:j})-1]\mathrm{\Phi}(v_{i}),\\
&\frac{\partial\mathcal{O}_{ij}^{s}}{W^{out,s}_{:k}} = \sigma(\mathrm{\Phi}(v_{i})\cdot W^{out,s}_{:k})\mathrm{\Phi}(v_{i}),\;\mathrm{for}\;v_{k}\in\mathcal{V}_{neg}.\nonumber
\end{aligned}
\end{equation}Similarly, after a node attribute co-occurrence pair $(v_{i},a_{j})$ is sampled, with negative sampling~\cite{gutmann2012noise}, the partial objective $\mathcal{O}^{a}_{ij}$ in Eq. (\ref{attribute_partial_obj}) is approximated as

\begin{equation}
\mathcal{O}^{a}_{ij} = -\log\sigma(\mathrm{\Phi}(v_{i})\cdot W_{:j}^{out,a})
-\sum_{k:v_{k}\in\mathcal{A}_{neg}}\log\sigma(-\mathrm{\Phi}(v_{i})\cdot W_{:k}^{out,a}),
\end{equation}where $\mathcal{A}_{neg}$ is the set of sampled negative attributes. We then update the parameters with gradient descent

\begin{equation}\label{update_attribute}
\begin{aligned}
&W^{in}_{i:} = W^{in}_{i:}  - \eta \frac{\partial\mathcal{O}^{a}_{ij}}{\partial W^{in}_{i:}},\\
&W^{out,a}_{:j} = W^{out,a}_{:j}  - \eta \frac{\partial\mathcal{O}^{a}_{ij}}{\partial W^{out,a}_{:j}},\\
&W^{out,a}_{:k} = W^{out,a}_{:k}  - \eta \frac{\partial\mathcal{O}^{a}_{ij}}{\partial W^{out,a}_{:k}},\; \mathrm{for} \; a_{k}\in\mathcal{A}_{neg}.\\
\end{aligned}
\end{equation}The gradients are calculated as

\begin{equation}
\begin{aligned}
&\frac{\partial\mathcal{O}_{ij}^{a}}{\partial W^{in}_{ir}} =
\beta [1-{\tanh(\beta W^{in}_{ir})}^{2}][\sigma(\mathrm{\Phi}(v_{i})\cdot W_{:j}^{out,a})-1]W_{rj}^{out,a}\\
&+\beta [1-{\tanh(\beta W^{in}_{ir})}^{2}] \sum_{k:v_{k}\in\mathcal{A}_{neg}}\sigma(\mathrm{\Phi}(v_{i})\cdot W_{:k}^{out,a})W_{rk}^{out,a},\\
&\frac{\partial\mathcal{O}_{ij}^{a}}{W^{out,a}_{:j}} = [\sigma(\mathrm{\Phi}(v_{i})\cdot W^{out,a}_{:j})-1]\mathrm{\Phi}(v_{i}),\\
&\frac{\partial\mathcal{O}_{ij}^{a}}{W^{out,a}_{:k}} = \sigma(\mathrm{\Phi}(v_{i})\cdot W^{out,a}_{:k})\mathrm{\Phi}(v_{i}),\;\mathrm{for}\;a_{k}\in\mathcal{A}_{neg}.\nonumber
\end{aligned}
\end{equation}After the parameters are learned, for node $v_{i}\in\mathcal{V}$, we construct its embedding $\mathrm{\Phi}(v_{i})$ as
\begin{equation} \label{rep_final}
\mathrm{\Phi}(v_{i})_{r}=
\left\lbrace
\begin{aligned}
&+1,\quad\mathrm{if}\; \tanh(\beta W^{in}_{ir}) \geqslant 0,\\
&-1,\quad\mathrm{if}\; \tanh(\beta W^{in}_{ir}) <0.
\end{aligned}
\right.
\end{equation}To obtain binary codes for efficient Hamming distance calculation, we store the $-1$ value of $\mathrm{\Phi}(v_{i})_{r}$ as 0 instead. 

\begin{algorithm}[t]
	\caption{BinaryNE: Binary Network Embedding}
	\begin{small}
		\label{alg:BinaryNE}
		\begin{algorithmic}[1]
			\REQUIRE ~~\\
			A given network $G=(\mathcal{V},\mathcal{E},\mathcal{A},X)$;
			\ENSURE ~~\\
			Binary node embedding $\mathrm{\Phi}(\cdot)$ for each $v_{i}\in\mathcal{V}$;
			\STATE $\mathbb{S}$ $\leftarrow$ generate a set of random walks on $G$;
			\STATE $n(v_i,v_j)$ $\leftarrow$ count the frequency of node context pairs ($v_{i},v_{j})$ in $\mathbb{S}$;
			\STATE $(W^{in}, W^{out,s}, W^{out,a})$ $\leftarrow$ initialization;
			\REPEAT
			\STATE draw a random number $\delta \in(0,1)$;
			\IF{$\delta\leqslant 0.5$}
			\STATE{$(v_i,v_j) \leftarrow$ sample a node context pair according to the distribution of $n(v_i,v_j)$;}
			\STATE{$\mathcal{V}_{neg}\leftarrow$ draw $K$ negative nodes;}
			\STATE{$(W^{in},W^{out,s}) \leftarrow$ update parameters with $(v_i,v_j,\mathcal{V}_{neg})$ according to Eq.~(\ref{update_structure});}
			\ELSE
			\STATE{$(v_i,a_j) \leftarrow$ sample a node attribute pair according to the distribution of $X_{ij}$;}
			\STATE{$\mathcal{A}_{neg}\leftarrow$ draw $K$ negative attributes;}
			\STATE{$(W^{in},W^{out,a}) \leftarrow$ update parameters with $(v_i,a_j,\mathcal{A}_{neg})$ according to Eq.~(\ref{update_attribute});}
			\ENDIF
			\UNTIL {maximum number of iterations expire;}\label{code:iteration}
			\STATE construct node embedding $\Phi(\cdot)$ with $W^{in}$ and Eq. (\ref{rep_final}); 
			\STATE \textbf{return} $\mathrm{\Phi}(\cdot)$;
		\end{algorithmic}
	\end{small}
\end{algorithm}

Algorithm~\ref{alg:BinaryNE} provides the pseudocode of the proposed BinaryNE algorithm. At Step 1, a set of random walks with length $L$ are generated by starting random walks at each node $v_{i}\in\mathcal{V}$ for $\gamma$ times. At Step 2, on the generated random walks, with $t$ window size, BinaryNE collects node context pairs $(v_{i},v_{j})$ and counts their occurrence frequencies $n(v_{i},v_{j})$. At Step 3, $W^{in}$ is initialized with random numbers sampled from a uniform distribution in the range of $[-\frac{1}{2d},\frac{1}{2d}]$, and $W^{out,s}$ and $W^{out,a}$ are initialized with zero. At Step 4-15, the parameters are updated with stochastic gradient descent. Each iteration starts from drawing a random switch variable $\delta\in(0,1)$ to determine which partial objective to be optimized. To optimize the structure preserving partial objective, BinaryNE randomly draws a node context pair $(v_{i},v_{j})$ according to the distribution of $n(v_{i},v_{j})$, and draws $K$ negative nodes, forming $\mathcal{V}_{neg}$, then updates the parameters with Eq. (\ref{update_structure}). To optimize the attribute preserving objective, BinaryNE draws a node attribute co-occurrence pair $(v_{i},a_{j})$ according the distribution of $X_{ij}$ and draws a negative attribute set $\mathcal{A}_{neg}$ with size $K$, then updates the parameters with Eq. (\ref{update_attribute}). For efficient node context pair and node attribute pair sampling, BinaryNE adopts the alias table~\cite{li2014reducing} method, which takes only $O(1)$ time at each sampling. Finally, BinaryNE constructs binary node representations $\mathrm{\Phi}(\cdot)$ with $W^{in}$ \jy{according to} Eq. (\ref{rep_final}).

The time complexity of BinaryNE is determined by only the dimension of node embeddings $d$ and the maximum number of iterations. The scale of the maximum number of iterations is $O(\max(nnz(X),|\mathcal{V}|))$, where $nnz(X)$ is the number of non-zero entries of $X$, and $|\mathcal{V}|$ is the scale of node context pairs collected via random walks. BinaryNE has a time complexity of $O(d\cdot\max(nnz(X),|\mathcal{V}|))$, which guarantees its ability to scale up to large-scale graphs.

\section{Experiments}
In this section, we conduct experiments on six real-world networks to evaluate the effectiveness of binary node representations learned by BinaryNE for node similarity search, including search precision, response time, and memory usage, \jy{as well as on node classification and node clustering. The detailed experimental settings are given in the Appendix.}

\subsection{Datasets}
Six real-world networks are used in the experiments, with the details as follows:

\begin{itemize}
	\item \textbf{Cora}\footnote{https://linqs.soe.ucsc.edu/data\label{fn:linqs}}~\cite{getoor2003link}. The Cora network is composed of 2,708 machine learning publications and their citation relationships. Theses publications are categorized into seven groups. Each publication is represented by a 1,433-dimensional binary vector, with each dimension denoting the presence/absence of the corresponding word. 
	\item \textbf{Citeseer}\footnotemark[\value{footnote}]~\cite{getoor2003link}. Citeseer is another citation network with 3,312 papers and 4,732 citation relations. There are 6 classes among papers. According to the occurrence of the corresponding word, each paper is described by a 3,703-dimensional binary vector.
	\item \textbf{BlogCatalog}\footnote{https://www4.comp.polyu.edu.hk/\textasciitilde xiaohuang/Code.html\label{fn:tamu}}~\cite{huang2017accelerated}. The BlogCatalog network is an online social network formed by BlogCatalog, a blogger community. The BlogCatalog network contains 5,196 users and 171,743 follower-followee relations. Users' groups are defined as the categories of their blogs. The keywords of users' blogs are used to construct users' feature vectors. Here, binary feature vectors are constructed, with only the keyword occurrence state concerned.
	\item \textbf{Flickr}\footnotemark[\value{footnote}]~\cite{huang2017accelerated}. Flickr is an online photo sharing platform. The Flickr network includes 7,575 users and 239,738 follower-followee relations. These users join in 9 predefined groups. Users' features are described by the tags of their images. Each user is represented by 12,047-dimensional binary vector, according to the occurrence/absence of the corresponding tag.
	\item \textbf{DBLP(Subgraph)} and \textbf{DBLP(Full)}. The DBLP(Full) network is formed by the papers, paper titles, and paper citations of the DBLP bibliographic network\footnote{https://aminer.org/citation (Version 3 is used)}~\cite{tang2008arnetminer}. In DBLP(Full), there are in total 1,632,442 papers and 2,327,450 citations. The DBLP(Subgraph) is a subgraph of the DBLP(Full) network, constructed by papers from four research areas: \textit{Database, Data Mining, Artificial Intelligence, and Computer Version}, which also act as paper labels.  DBLP(Subgraph) contains 18,448 papers and 45,611 citation relations. For DBLP(Full) and DBLP(Subgraph), papers' titles are used to construct binary bag-of-words feature vectors.
\end{itemize}

\begin{table}[t]
	\begin{center}
		\tabcolsep 3pt
		\caption{Summary of six real-world networks}
		\renewcommand{\arraystretch}{1.25}
		\scalebox{0.8}{
			\begin{tabular}{cccccc}
				\hline
				& $|\mathcal{V}|$ & $|\mathcal{E}|$ & $|\mathcal{A}|$ & $nnz(X)$ & \# of Class\\\hline
				Cora & 2,708 & 5,278 & 1,433 & 49,216 & 7\\
				Citeseer & 3,312 & 4,732 & 3,703 & 105,165 & 6 \\
				BlogCatalog	& 5,196 & 171,743 & 8,189 & 369,435 & 6 \\
				Flickr	& 7,575 & 239,738 & 12,047 & 182,517 & 9 \\
				DBLP(Subgraph) & 18,448 & 45,611 & 2,476 & 103,130 & 4\\
				DBLP(Full) & 1,632,442 & 2,327,450 & 154,309 & 10,413,178 & N/A\\\hline
		\end{tabular}}
		\label{dataset}
	\end{center}
\end{table}

Table \ref{dataset} summarizes the statistics of the networks. For each network, the direction of links is ignored. \textbf{Cora}, \textbf{Citeseer}, \textbf{BlogCatalog}, \textbf{Flickr}, and \textbf{DBLP(Subgraph)} are used to evaluate the performance of the binary node representations learned by BinaryNE on node similarity search, including search precision, query time, and memory usage, as well as on \jy{node classification and node clustering}. \textbf{DBLP(Full)} is used to investigate the scalability of node similarity search with BinaryNE binary codes.

\subsection{Baseline Methods}
BinaryNE is compared with two groups of the state-of-the-art methods:

\begin{itemize}
	\item \textbf{Continuous embeddings measured by Euclidean distance}: 
	\begin{itemize}
	\item \textbf{DeepWalk}/\textbf{node2vec}~\cite{perozzi2014deepwalk,grover2016node2vec} preserves the similarity between nodes sharing similar context in random walks. node2vec is equivalent to DeepWalk with the default parameter setting $p=q=1$.
	
	\item\textbf{LINE1}~\cite{tang2015line} denotes the version of LINE that captures the first-order proximity. 
	
	\item \textbf{LINE2}~\cite{tang2015line} represents the version of LINE that models the second-order proximity.
	
	\item \textbf{SDNE}~\cite{wang2016structural} learns deep non-linear node representations via a semi-supervised deep autoencoder.
	
	\item \textbf{TADW}~\cite{yang2015network} learns node embeddings that capture both network structure and node content similarity via inductive matrix factorization~\cite{natarajan2014inductive}. 
	
	\item \textbf{UPP-SNE}~\cite{zhang2017user} performs a non-linear mapping on node content features to learn node embeddings that preserve both network structure and node content features. 
	
	\item \textbf{MVC-DNE}~\cite{yang2017properties} fuses network structure and node content features into node embeddings through deep cross-view learning.
	
	\item \textbf{SINE}~\cite{zhang2018sine} learns node representations by using node representations to simultaneously predict context nodes and node attributes.
	
	\item \textbf{Feature}. Node raw content feature is also used as a baseline for similarity search. For each node $v_{i}\in\mathcal{V}$, its feature vector is $X_{i:}$, with $X_{i:}$ being the $i$-th row of $X$.
	\end{itemize}
	
	\item \textbf{Discrete embeddings measured by Hamming distance}:
	\begin{itemize}
	\item \textbf{Quantized Continuous Embeddings}. To obtain binary node representations, a naive way is to quantize the continuous node embeddings into binary codes.  As a baseline, we binarize the continuous embeddings learned by above baseline methods with \dk{the state-of-the-art hash learning algorithm: Iterative Quantization~\cite{gong2012iterative}}, and denote these methods as \textbf{DeepWalk+Q}, \textbf{LINE1+Q}, \textbf{LINE2+Q}, \textbf{SDNE+Q}, \textbf{TADW+Q}, \textbf{UPP-SNE+Q}, \textbf{MVC-DNE+Q}, \textbf{SINE+Q}, and \textbf{Feature+Q}, respectively. 
		
	\item \textbf{BANE}~\cite{yang2018binarized} learns binary node representations by performing binary factorization on the Weisfeiler-Lehman proximity matrix~\cite{shervashidze2011weisfeiler} that carries both network structure and node content information.
	
	\item \dk{\textbf{KD-coding}~\cite{chen2018learning} learns $K$-way $D$-dimensional discrete node representations from the continuous node representations learned by DeepWalk. To obtain 128-dimensional binary codes, we set $K=2$ and $D=128$, respectively.}
		
	\item \dk{\textbf{DNE}~\cite{xiaobo2018discrete} learns binary node representations through discrete matrix factorization in a supervised manner. To enable DNE to work under our unsupervised setting, we set the weight of its supervised learning component to 0 in the optimization objective.}

	\item \textbf{NetHash}~\cite{wu2018efficient} generates discrete node embeddings by randomly sampling the union of neighboring node attributes. As the learned discrete node representations do not take binary values, the Hamming distance cannot be efficiently calculated with bit-wise operations.

	\item \textbf{BinaryNE\_S} is the ablated version of BinaryNE that uses only network structure to learn binary node representations.
	
	\item \textbf{BinaryNE\_C} is the ablated version of BinaryNE that uses only node content attributes to learn binary node codes.
	
	\item \textbf{BinaryNE\_S+C} concatenates the binary codes learned via BinaryNE\_S and BinaryNE\_C, which is another ablated version of BinaryNE that respectively encodes network structure and node content attributes into seperated dimensions of binary node representations. The binary codes learned via BinaryNE\_S and BinaryNE\_C have half dimensions of BinaryNE's binary codes for fair comparisons. 
	\end{itemize}
\end{itemize}

\subsection{Evaluation Metrics}

For all methods, we use the entire network to learn continuous or binary node representations for all nodes and evaluate the node similarity search performance by searching similar nodes for every node in the network from the remaining nodes.

For each node in a network, we in turn query its top-$K$ similar nodes. $K$ is set to 100, 200, and 500, respectively. We adopt averaged \textit{precision} and \textit{MAP (Mean Averaged Precision)} as evaluation metrics. 

For querying nodes similar to node $v_{i}$, the $precision@K(v_{i})$ is defined as
\begin{equation}
precision@K(v_{i}) = \frac{|\{v_{j}|rank(v_{j})\leq K, C(v_{i})=C(v_{j})\}|}{K},\nonumber
\end{equation}where $rank(v_{j})$ is the position of $v_{j}$ in the rank list of nodes similar to $v_{i}$. $C(v_{i})=C(v_{j})$ indicates that node $v_{i}$ and $v_{j}$ have the same class label, with $C(\cdot)$ denoting node class label. As we in turn take all nodes in $\mathcal{V}$ as query nodes, we report the averaged $precision@K$ as final results.

MAP (Mean Average Precision) is an information retrieval metric with good discrimination and stability. Different from precision, MAP takes into account the order in which relevant nodes are placed in the returned rank list. When we vary the query node $v_{i}$ over $\mathcal{V}$, the MAP value is calculated as
\begin{equation}
\begin{aligned}
&AP@K(v_{i}) = \frac{\sum_{k=1}^{K}precision@k(v_{i})\cdot relavant@k(v_{i})}{|\{v_{j}|C(v_{j})=C(v_{i}),v_{j}\in\mathcal{V}\}|},\\
&MAP@K = \frac{\sum_{i=1}^{|\mathcal{V}|}AP@K(v_{i})}{|\mathcal{V}|},\nonumber
\end{aligned}
\end{equation}where $relavant@k(v_{i})$ is an indicator function equaling 1 if the $k$-th retrieved node is relevant to $v_i$ and 0 otherwise.

\subsection{Similarity Search Results}
Tables~\ref{Res_Cora}-\ref{Res_DBLP(Subgraph)} give similarity search results on Cora, Citeseer, BlogCatalog, Flickr, and DBLP(Subgraph). For query time, we only consider the time consumed by calculating the distance between the query node and all remaining nodes, which contributes to the main computational overhead of similarity search, and report the time averaged over all query nodes (in milliseconds). We provide the search speedup of BinaryNE compared with baselines. We also perform paired t-tests between the search precisions delivered by BinaryNE and baseline methods, where we use $\bullet$ ($\circ$) to indicate that BinaryNE is significantly better (worse) than the compared baseline methods at $95\%$ significance level. For all methods, the best and second best performers are highlighted by \textbf{bold} and \underline{underline}, respectively.

\begin{table*}[t]
	\centering
	\scriptsize
	\tabcolsep 2pt
	\caption{Similarity search results on Cora}
	\renewcommand{\arraystretch}{1}
	\begin{tabular}{cccccccccc} 
		\hline
		Metric & Method & precision@100 & MAP@100 & precision@200 & MAP@200 & precision@500 & MAP@500  & Query time (ms) & Speedup\\\hline
		\multirow{9}{*}{Euclidean}
		& DeepWalk & 0.5621 $\bullet$ & \underline{0.1149} $\circ$ & 0.4782 $\bullet$ & \underline{0.1771} $\circ$ & 0.3464 $\bullet$ & 0.2616 $\bullet$ & 1.62 & 32.5 $\times$ \\
		& LINE1 & 0.4042 $\bullet$ & 0.0678 $\bullet$ & 0.3002 $\bullet$ & 0.0828 $\bullet$ & 0.2270 $\bullet$ & 0.1110 $\bullet$ & 1.62 & 32.5 $\times$ \\
		& LINE2 & 0.3425 $\bullet$ & 0.0482 $\bullet$ & 0.2868 $\bullet$ & 0.0645 $\bullet$ & 0.2411 $\bullet$ & 0.0993 $\bullet$ & 1.90 & 37.9 $\times$ \\
		& SDNE & 0.3595 $\bullet$ & 0.0565 $\bullet$ & 0.2901 $\bullet$ & 0.0730 $\bullet$ & 0.2370 $\bullet$ & 0.1058 $\bullet$ & 1.62 & 32.5 $\times$ \\
		& TADW & 0.4508 $\bullet$ & 0.0871 $\bullet$ & 0.3621 $\bullet$ & 0.1227 $\bullet$ & 0.2607 $\bullet$ & 0.1682 $\bullet$ & 1.62 & 32.5 $\times$ \\
		& UPP-SNE & \textbf{0.6032} $\circ$ & \textbf{0.1204} $\circ$ & \textbf{0.5242} $\circ$ & \textbf{0.1890} $\circ$ & \underline{0.3990} $\bullet$ & \textbf{0.2947} $\circ$ & 1.62 & 32.5 $\times$ \\
		& MVC-DNE & 0.3559 $\bullet$ & 0.0408 $\bullet$ & 0.3190 $\bullet$ & 0.0626 $\bullet$ & 0.2738 $\bullet$ & 0.1094 $\bullet$ & 1.62 & 32.5 $\times$ \\
		& SINE & 0.4309 $\bullet$ & 0.0701 $\bullet$ & 0.3703 $\bullet$ & 0.1026 $\bullet$ & 0.2968 $\bullet$ & 0.1611 $\bullet$ & 1.62 & 32.5 $\times$ \\
		& Feature & 0.2240 $\bullet$ & 0.0166 $\bullet$ & 0.2060 $\bullet$ & 0.0252 $\bullet$ & 0.2189 $\bullet$ & 0.0551 $\bullet$ & 19.23 & 384.5 $\times$ \\\hline
		\multirow{17}{*}{Hamming} 
		& DeepWalk+Q & \underline{0.5748} $\;\;$ & 0.1139 $\circ$ & 0.4912 $\bullet$ & 0.1749 $\circ$ & 0.3703 $\bullet$ & 0.2666 $\bullet$ & 0.05 & 1.0 $\times$ \\
		& LINE1+Q & 0.3916 $\bullet$ & 0.0688 $\bullet$ & 0.2922 $\bullet$ & 0.0833 $\bullet$ & 0.2160 $\bullet$ & 0.1087 $\bullet$ & 0.05 & 1.0 $\times$ \\
		& LINE2+Q & 0.3880 $\bullet$ & 0.0606 $\bullet$ & 0.3203 $\bullet$ & 0.0834 $\bullet$ & 0.2516 $\bullet$ & 0.1219 $\bullet$ & 0.05 & 1.0 $\times$ \\
		& SDNE+Q & 0.4522 $\bullet$ & 0.0838 $\bullet$ & 0.3635 $\bullet$ & 0.1146 $\bullet$ & 0.2666 $\bullet$ & 0.1552 $\bullet$ & 0.05 & 1.0 $\times$ \\
		& TADW+Q & 0.5621 $\bullet$ & 0.1138 $\circ$ & 0.4570 $\bullet$ & 0.1623 $\bullet$ & 0.3310 $\bullet$ & 0.2282 $\bullet$ & 0.06 & 1.2 $\times$ \\
		& UPP-SNE+Q & 0.5520 $\bullet$ & 0.1031 $\bullet$ & 0.4732 $\bullet$ & 0.1572 $\bullet$ & 0.3606 $\bullet$ & 0.2421 $\bullet$ & 0.05 & 1.0 $\times$ \\
		& MVC-DNE+Q & 0.3323 $\bullet$ & 0.0348 $\bullet$ & 0.3013 $\bullet$ & 0.0546 $\bullet$ & 0.2609 $\bullet$ & 0.0976 $\bullet$ & 0.07 & 1.4 $\times$ \\
		& SINE+Q & 0.4564 $\bullet$ & 0.0715 $\bullet$ & 0.3909 $\bullet$ & 0.1055 $\bullet$ & 0.3114 $\bullet$ & 0.1667 $\bullet$ & 0.06 & 1.2 $\times$ \\
		& Feature+Q & 0.3701 $\bullet$ & 0.0459 $\bullet$ & 0.3285 $\bullet$ & 0.0702 $\bullet$ & 0.2740 $\bullet$ & 0.1179 $\bullet$ & 0.06 & 1.2 $\times$ \\
		& BANE & 0.2225 $\bullet$ & 0.0168 $\bullet$ & 0.2086 $\bullet$ & 0.0260 $\bullet$ & 0.1960 $\bullet$ & 0.0499 $\bullet$ & 0.05 & 1.0 $\times$\\
		& DNE & 0.4781 $\bullet$ & 0.0969 $\bullet$ & 0.3906 $\bullet$ & 0.1397 $\bullet$ & 0.2838 $\bullet$ & 0.1886 $\bullet$ & 0.05 & 1.0 $\times$ \\
		& KD-coding & 0.5092 $\bullet$ & 0.0874 $\bullet$ & 0.4640 $\bullet$ & 0.1460 $\bullet$ & 0.3795 $\bullet$ & 0.2570 $\bullet$ & 0.05 & 1.0 $\times$ \\
		& NetHash & 0.4546 $\bullet$ & 0.0757 $\bullet$ & 0.3852 $\bullet$  & 0.1097 $\bullet$  & 0.2993 $\bullet$  & 0.1656 $\bullet$  & 1.17 & 23.4 $\times$\\
		& BinaryNE\_S & 0.5598 $\bullet$ & 0.1072 $\circ$ & 0.4916 $\bullet$ & 0.1693 $\;\;$ & 0.3748 $\bullet$ & 0.2603 $\bullet$ & 0.05 & 1.0 $\times$ \\
		& BinaryNE\_C & 0.3553 $\bullet$ & 0.0391 $\bullet$ & 0.3217 $\bullet$ & 0.0623 $\bullet$ & 0.2768 $\bullet$ & 0.1114 $\bullet$ & 0.05 & 1.0 $\times$ \\
		& BinaryNE\_S+C & 0.5483 $\bullet$ & 0.1009 $\bullet$ & 0.4811 $\bullet$ & 0.1592 $\bullet$ & 0.3738 $\bullet$ & 0.2543 $\bullet$ & 0.05 & 1.0 $\times$ \\
		& BinaryNE & 0.5723 $\;\;$ & 0.1050 $\;\;$ & \underline{0.5091} $\;\;$ & 0.1694 $\;\;$ & \textbf{0.4070} $\;\;$ & \underline{0.2848} $\;\;$ & 0.05 & \\
		\hline
	\end{tabular}
	\label{Res_Cora}
\end{table*}

\begin{table*}[t]
	\centering
	\scriptsize
	\tabcolsep 2pt
	\caption{Similarity search results on Citeseer}
	\renewcommand{\arraystretch}{1}
	\begin{tabular}{cccccccccc} 
		\hline
		Metric&Method & precision@100 & MAP@100 & precision@200 & MAP@200 & precision@500 & MAP@500 & Query time (ms) & Speedup\\\hline
		\multirow{9}{*}{Euclidean}
		& DeepWalk & 0.3806 $\bullet$ & 0.0419 $\bullet$ & 0.3165 $\bullet$ & 0.0597 $\bullet$ & 0.2534 $\bullet$ & 0.0923 $\bullet$ & 1.99 & 33.1 $\times$ \\
		& LINE1 & 0.2875 $\bullet$ & 0.0282 $\bullet$ & 0.2360 $\bullet$ & 0.0372 $\bullet$ & 0.1997 $\bullet$ & 0.0573 $\bullet$ & 1.99 & 33.1 $\times$ \\
		& LINE2 & 0.2493 $\bullet$ & 0.0210 $\bullet$ & 0.2171 $\bullet$ & 0.0286 $\bullet$ & 0.1943 $\bullet$ & 0.0473 $\bullet$ & 1.99 & 33.1 $\times$ \\
		& SDNE & 0.2436 $\bullet$ & 0.0210 $\bullet$ & 0.2091 $\bullet$ & 0.0282 $\bullet$ & 0.1866 $\bullet$ & 0.0469 $\bullet$ & 1.99 & 33.1 $\times$ \\
		& TADW & 0.3728 $\bullet$ & 0.0381 $\bullet$ & 0.3213 $\bullet$ & 0.0572 $\bullet$ & 0.2609 $\bullet$ & 0.0944 $\bullet$ & 1.99 & 33.1 $\times$ \\
		& UPP-SNE & \textbf{0.4951} $\circ$ & \textbf{0.0591} $\circ$ & \textbf{0.4504} $\circ$ & \textbf{0.1000} $\circ$ & \underline{0.3714} $\bullet$ & \underline{0.1801} $\;\;$ & 1.99 & 33.1 $\times$ \\
		& MVC-DNE & 0.3478 $\bullet$ & 0.0293 $\bullet$ & 0.3143 $\bullet$ & 0.0465 $\bullet$ & 0.2724 $\bullet$ & 0.0847 $\bullet$ & 1.99 & 33.1 $\times$ \\
		& SINE & 0.3831 $\bullet$ & 0.0376 $\bullet$ & 0.3418 $\bullet$ & 0.0590 $\bullet$ & 0.2882 $\bullet$ & 0.1023 $\bullet$ & 1.99 & 33.1 $\times$ \\
		& feature & 0.2532 $\bullet$ & 0.0140 $\bullet$ & 0.2471 $\bullet$ & 0.0249 $\bullet$ & 0.2320 $\bullet$ & 0.0530 $\bullet$ & 54.65 & 910.8 $\times$ \\
		\hline
		\multirow{17}{*}{Hamming}
		& DeepWalk+Q & 0.3854 $\bullet$ & 0.0417 $\bullet$ & 0.3384 $\bullet$ & 0.0647 $\bullet$ & 0.2771 $\bullet$ & 0.1073 $\bullet$ & 0.07 & 1.2 $\times$ \\
		& LINE1+Q & 0.2785 $\bullet$ & 0.0266 $\bullet$ & 0.2317 $\bullet$ & 0.0350 $\bullet$ & 0.1967 $\bullet$ & 0.0539 $\bullet$ & 0.06 & 1.0 $\times$ \\
		& LINE2+Q & 0.2777 $\bullet$ & 0.0244 $\bullet$ & 0.2384 $\bullet$ & 0.0339 $\bullet$ & 0.2063 $\bullet$ & 0.0553 $\bullet$ & 0.07 & 1.2 $\times$ \\
		& SDNE+Q & 0.3232 $\bullet$ & 0.0327 $\bullet$ & 0.2724 $\bullet$ & 0.0462 $\bullet$ & 0.2249 $\bullet$ & 0.0722 $\bullet$ & 0.06 & 1.0 $\times$ \\
		& TADW+Q & 0.4212 $\bullet$ & 0.0461 $\bullet$ & 0.3605 $\bullet$ & 0.0692 $\bullet$ & 0.2877 $\bullet$ & 0.1112 $\bullet$ & 0.07 & 1.2 $\times$ \\
		& UPP-SNE+Q & 0.4768 $\bullet$ & 0.0560 $\bullet$ & 0.4332 $\bullet$ & 0.0942 $\bullet$ & 0.3578 $\bullet$ & 0.1687 $\bullet$ & 0.07 & 1.2 $\times$ \\
		& MVC-DNE+Q & 0.3165 $\bullet$ & 0.0239 $\bullet$ & 0.2900 $\bullet$ & 0.0386 $\bullet$ & 0.2574 $\bullet$ & 0.0730 $\bullet$ & 0.07 & 1.2 $\times$ \\
		& SINE+Q & 0.3435 $\bullet$ & 0.0292 $\bullet$ & 0.3098 $\bullet$ & 0.0460 $\bullet$ & 0.2682 $\bullet$ & 0.0833 $\bullet$ & 0.07 & 1.2 $\times$ \\
		& Feature+Q & 0.3701 $\bullet$ & 0.0331 $\bullet$ & 0.3343 $\bullet$ & 0.0532 $\bullet$ & 0.2841 $\bullet$ & 0.0952 $\bullet$ & 0.07 & 1.2 $\times$ \\
		& BANE & 0.2300 $\bullet$ & 0.0136 $\bullet$ & 0.2162 $\bullet$ & 0.0219 $\bullet$ & 0.2007 $\bullet$ & 0.0426 $\bullet$ & 0.06 & 1.0 $\times$\\
		& DNE & 0.3030 $\bullet$ & 0.0294 $\bullet$ & 0.2675 $\bullet$ & 0.0445 $\bullet$ & 0.2271 $\bullet$ & 0.0732 $\bullet$ & 0.07 & 1.2 $\times$ \\
		& KD-coding & 0.2572 $\bullet$ & 0.0199 $\bullet$ & 0.2496 $\bullet$ & 0.0397 $\bullet$ & 0.2538 $\bullet$ & 0.1033 $\bullet$ & 0.06 & 1.0 $\times$ \\
		& NetHash & 0.3866 $\bullet$ & 0.0378 $\bullet$ & 0.3417 $\bullet$ & 0.0583 $\bullet$  & 0.2851 $\bullet$ & 0.0999 $\bullet$ & 1.35 & 22.5 $\times$\\
		& BinaryNE\_S & 0.3975 $\bullet$ & 0.0440 $\bullet$ & 0.3641 $\bullet$ & 0.0741 $\bullet$ & 0.3072 $\bullet$ & 0.1350 $\bullet$ & 0.06 & 1.0 $\times$ \\
		& BinaryNE\_C & 0.3528 $\bullet$ & 0.0286 $\bullet$ & 0.3241 $\bullet$ & 0.0473 $\bullet$ & 0.2847 $\bullet$ & 0.0896 $\bullet$ & 0.06 & 1.0 $\times$ \\
		& BinaryNE\_S+C & 0.4369 $\bullet$ & 0.0485 $\bullet$ & 0.4003 $\bullet$ & 0.0820 $\bullet$ & 0.3393 $\bullet$ & 0.1507 $\bullet$ & 0.06 & 1.0 $\times$ \\
		& BinaryNE & \underline{0.4846} $\;\;$ & \underline{0.0572} $\;\;$ & \underline{0.4454} $\;\;$ & \underline{0.0981} $\;\;$ & \textbf{0.3758} $\;\;$ & \textbf{0.1814} $\;\;$ & 0.06 & \\
		\hline
	\end{tabular}
	\label{Res_Citeseer}
\end{table*}

\begin{table*}[t]
	\centering
	\scriptsize
	\tabcolsep 2pt
	\caption{Similarity search results on BlogCatalog}
	\renewcommand{\arraystretch}{1}
	\begin{tabular}{ccccccccccc} 
		\hline
		Metric & Method & precision@100 & MAP@100 & precision@200 & MAP@200 & precision@500 & MAP@500 & Query time (ms) & Speedup\\\hline
		\multirow{9}{*}{Euclidean}
		& DeepWalk & 0.4349 $\bullet$ & 0.0327 $\bullet$ & 0.3818 $\bullet$ & 0.0521 $\bullet$ & 0.3005 $\bullet$ & 0.0859 $\bullet$ & 3.12 & 31.2 $\times$ \\
		& LINE1 & 0.3850 $\bullet$ & 0.0275 $\bullet$ & 0.3160 $\bullet$ & 0.0411 $\bullet$ & 0.2375 $\bullet$ & 0.0616 $\bullet$ & 3.12 & 31.2 $\times$ \\
		& LINE2 & 0.2412 $\bullet$ & 0.0140 $\bullet$ & 0.2239 $\bullet$ & 0.0226 $\bullet$ & 0.2029 $\bullet$ & 0.0420 $\bullet$ & 2.60 & 26.0 $\times$ \\
		& SDNE & 0.3145 $\bullet$ & 0.0201 $\bullet$ & 0.2795 $\bullet$ & 0.0316 $\bullet$ & 0.2366 $\bullet$ & 0.0548 $\bullet$ & 3.12 & 31.2 $\times$ \\
		& TADW & \underline{0.7431} $\circ$ & \underline{0.0758} $\circ$ & \underline{0.6923} $\circ$ & \underline{0.1373} $\circ$ & 0.5675 $\bullet$ & \underline{0.2621} $\circ$ & 2.60 & 26.0 $\times$ \\
		& UPP-SNE & 0.5173 $\bullet$ & 0.0417 $\bullet$ & 0.4735 $\bullet$ & 0.0707 $\bullet$ & 0.3931 $\bullet$ & 0.1281 $\bullet$ & 2.60 & 26.0 $\times$ \\
		& MVC-DNE & 0.5616 $\bullet$ & 0.0463 $\bullet$ & 0.5008 $\bullet$ & 0.0761 $\bullet$ & 0.4109 $\bullet$ & 0.1356 $\bullet$ & 3.12 & 31.2 $\times$ \\
		& SINE & 0.3502 $\bullet$ & 0.0216 $\bullet$ & 0.3067 $\bullet$ & 0.0326 $\bullet$ & 0.2599 $\bullet$ & 0.0567 $\bullet$ & 3.12 & 31.2 $\times$ \\
		& Feature & 0.2424 $\bullet$ & 0.0113 $\bullet$ & 0.2239 $\bullet$ & 0.0177 $\bullet$ & 0.2023 $\bullet$ & 0.0333 $\bullet$ & 184.98 & 1849.8 $\times$ \\
		\hline
		\multirow{17}{*}{Hamming} 
		& DeepWalk+Q & 0.3950 $\bullet$ & 0.0266 $\bullet$ & 0.3545 $\bullet$ & 0.0432 $\bullet$ & 0.2883 $\bullet$ & 0.0728 $\bullet$ & 0.11 & 1.1 $\times$ \\
		& LINE1+Q & 0.4021 $\bullet$ & 0.0275 $\bullet$ & 0.3529 $\bullet$ & 0.0436 $\bullet$ & 0.2779 $\bullet$ & 0.0701 $\bullet$ & 0.11 & 1.1 $\times$ \\
		& LINE2+Q & 0.2958 $\bullet$ & 0.0146 $\bullet$ & 0.2720 $\bullet$ & 0.0236 $\bullet$ & 0.2402 $\bullet$ & 0.0440 $\bullet$ & 0.10 & 1.0 $\times$ \\
		& SDNE+Q & 0.3538 $\bullet$ & 0.0219 $\bullet$ & 0.3127 $\bullet$ & 0.0342 $\bullet$ & 0.2607 $\bullet$ & 0.0578 $\bullet$ & 0.12 & 1.2 $\times$ \\
		& TADW+Q & \textbf{0.7448} $\circ$ & \textbf{0.0776} $\circ$ & \textbf{0.7081} $\circ$ & \textbf{0.1453} $\circ$ & \textbf{0.6147} $\circ$ & \textbf{0.3028} $\circ$ & 0.10 & 1.0 $\times$ \\
		& UPP-SNE+Q & 0.4830 $\bullet$ & 0.0372 $\bullet$ & 0.4416 $\bullet$ & 0.0628 $\bullet$ & 0.3655 $\bullet$ & 0.1123 $\bullet$ & 0.10 & 1.0 $\times$ \\
		& MVC-DNE+Q & 0.5117 $\bullet$ & 0.0393 $\bullet$ & 0.4568 $\bullet$ & 0.0642 $\bullet$ & 0.3804 $\bullet$ & 0.1154 $\bullet$ & 0.11 & 1.1 $\times$ \\
		& SINE+Q & 0.3745 $\bullet$ & 0.0225 $\bullet$ & 0.3363 $\bullet$ & 0.0356 $\bullet$ & 0.2903 $\bullet$ & 0.0650 $\bullet$ & 0.10 & 1.0 $\times$ \\
		& Feature+Q & 0.4921 $\bullet$ & 0.0396 $\bullet$ & 0.4473 $\bullet$ & 0.0666 $\bullet$ & 0.3798 $\bullet$ & 0.1244 $\bullet$ & 0.10 & 1.0 $\times$ \\
		& BANE & 0.4892 $\bullet$ & 0.0371 $\bullet$ & 0.4572 $\bullet$ & 0.0655 $\bullet$ & 0.4023 $\bullet$ & 0.1308 $\bullet$ & 0.11 & 1.1 $\times$\\
		& DNE & 0.1689 $\bullet$ & 0.0044 $\bullet$ & 0.1689 $\bullet$ & 0.0080 $\bullet$ & 0.1692 $\bullet$ & 0.0191 $\bullet$ & 0.09 & 0.9 $\times$ \\
		& KD-coding & 0.3754 $\bullet$ & 0.0242 $\bullet$ & 0.3455 $\bullet$ & 0.0409 $\bullet$ & 0.2887 $\bullet$ & 0.0724 $\bullet$ & 0.08 & 0.8 $\times$ \\
		& NetHash& 0.3811 $\bullet$ & 0.0246 $\bullet$ & 0.3388 $\bullet$ & 0.0385 $\bullet$ & 0.2882 $\bullet$ & 0.0684 $\bullet$ & 3.14 & 31.4 $\times$\\
		& BinaryNE\_S & 0.4729 $\bullet$ & 0.0352 $\bullet$ & 0.4347 $\bullet$ & 0.0604 $\bullet$ & 0.3672 $\bullet$ & 0.1126 $\bullet$ & 0.09 & 0.9 $\times$ \\
		& BinaryNE\_C & 0.5571 $\bullet$ & 0.0466 $\bullet$ & 0.5181 $\bullet$ & 0.0820 $\bullet$ & 0.4531 $\bullet$ & 0.1647 $\bullet$ & 0.10 & 1.0 $\times$ \\
		& BinaryNE\_S+C & 0.6500 $\bullet$ & 0.0601 $\bullet$ & 0.5846 $\bullet$ & 0.1024 $\bullet$ & 0.4769 $\bullet$ & 0.1870 $\bullet$ & 0.10 & 1.0 $\times$ \\
		& BinaryNE & 0.7081 $\;\;$ & 0.0687 $\;\;$ & 0.6637 $\;\;$ & 0.1239 $\;\;$ & \underline{0.5828} $\;\;$ & 0.2547 $\;\;$ & 0.10 & \\
		\hline
	\end{tabular}
	\label{Res_BlogCatalog}
\end{table*}

\begin{table*}[t]
	\centering
	\scriptsize
	\tabcolsep 2pt
	\caption{Similarity search results on Flickr}
	\renewcommand{\arraystretch}{1}
	\begin{tabular}{cccccccccc} 
		\hline
		Metric & Method & precision@100 & MAP@100 & precision@200 & MAP@200 & precision@500 & MAP@500 & Query time (ms) & Speedup\\\hline
		\multirow{9}{*}{Euclidean} 
		& DeepWalk & 0.2194 $\bullet$ & 0.0101 $\bullet$ & 0.2014 $\bullet$ & 0.0162 $\bullet$ & 0.1762 $\bullet$ & 0.0287 $\bullet$ & 4.54 & 32.5 $\times$ \\
		& LINE1 & 0.2318 $\bullet$ & 0.0125 $\bullet$ & 0.2075 $\bullet$ & 0.0198 $\bullet$ & 0.1747 $\bullet$ & 0.0333 $\bullet$ & 4.54 & 32.5 $\times$ \\
		& LINE2 & 0.1573 $\bullet$ & 0.0074 $\bullet$ & 0.1464 $\bullet$ & 0.0122 $\bullet$ & 0.1357 $\bullet$ & 0.0239 $\bullet$ & 3.79 & 27.1 $\times$ \\
		& SDNE & 0.1693 $\bullet$ & 0.0082 $\bullet$ & 0.1548 $\bullet$ & 0.0131 $\bullet$ & 0.1408 $\bullet$ & 0.0251 $\bullet$ & 3.79 & 27.1 $\times$ \\
		& TADW & 0.3644 $\bullet$ & 0.0278 $\bullet$ & 0.3104 $\bullet$ & 0.0421 $\bullet$ & 0.2368 $\bullet$ & 0.0649 $\bullet$ & 3.79 & 27.1 $\times$ \\
		& UPP-SNE & 0.3917 $\bullet$ & 0.0309 $\bullet$ & 0.3594 $\bullet$ & 0.0525 $\bullet$ & 0.3142 $\bullet$ & 0.0990 $\bullet$ & 4.54 & 32.5 $\times$ \\
		& MVC-DNE & 0.3047 $\bullet$ & 0.0176 $\bullet$ & 0.2723 $\bullet$ & 0.0275 $\bullet$ & 0.2329 $\bullet$ & 0.0489 $\bullet$ & 4.54 & 32.5 $\times$ \\
		& SINE & 0.3053 $\bullet$ & 0.0180 $\bullet$ & 0.2630 $\bullet$ & 0.0266 $\bullet$ & 0.2124 $\bullet$ & 0.0436 $\bullet$ & 3.79 & 27.1 $\times$ \\
		& Feature & 0.1379 $\bullet$ & 0.0055 $\bullet$ & 0.1275 $\bullet$ & 0.0082 $\bullet$ & 0.1190 $\bullet$ & 0.0152 $\bullet$ & 415.11 & 2965.1 $\times$ \\
		\hline
		\multirow{17}{*}{Hamming} 
		& DeepWalk+Q & 0.2348 $\bullet$ & 0.0119 $\bullet$ & 0.2155 $\bullet$ & 0.0193 $\bullet$ & 0.1855 $\bullet$ & 0.0334 $\bullet$ & 0.14 & 1.0 $\times$ \\
		& LINE1+Q & 0.2505 $\bullet$ & 0.0129 $\bullet$ & 0.2244 $\bullet$ & 0.0203 $\bullet$ & 0.1875 $\bullet$ & 0.0338 $\bullet$ & 0.16 & 1.1 $\times$ \\
		& LINE2+Q & 0.2135 $\bullet$ & 0.0083 $\bullet$ & 0.1974 $\bullet$ & 0.0135 $\bullet$ & 0.1764 $\bullet$ & 0.0253 $\bullet$ & 0.15 & 1.1 $\times$ \\
		& SDNE+Q & 0.2352 $\bullet$ & 0.0109 $\bullet$ & 0.2127 $\bullet$ & 0.0171 $\bullet$ & 0.1835 $\bullet$ & 0.0298 $\bullet$ & 0.15 & 1.1 $\times$ \\
		& TADW+Q & 0.3952 $\bullet$ & 0.0313 $\bullet$ & 0.3601 $\bullet$ & 0.0522 $\bullet$ & 0.3008 $\bullet$ & 0.0927 $\bullet$ & 0.15 & 1.1 $\times$ \\
		& UPP-SNE+Q & 0.4939 $\bullet$ & 0.0434 $\bullet$ & 0.4642 $\bullet$ & 0.0785 $\bullet$ & \underline{0.4028} $\bullet$ & \underline{0.1573} $\bullet$ & 0.16 & 1.1 $\times$ \\
		& MVC-DNE+Q & 0.3028 $\bullet$ & 0.0167 $\bullet$ & 0.2711 $\bullet$ & 0.0263 $\bullet$ & 0.2317 $\bullet$ & 0.0468 $\bullet$ & 0.14 & 1.0 $\times$ \\
		& SINE+Q & 0.3660 $\bullet$ & 0.0248 $\bullet$ & 0.3154 $\bullet$ & 0.0366 $\bullet$ & 0.2575 $\bullet$ & 0.0605 $\bullet$ & 0.15 & 1.1 $\times$ \\
		& Feature+Q & 0.4733 $\bullet$ & 0.0430 $\bullet$ & 0.4063 $\bullet$ & 0.0662 $\bullet$ & 0.3107 $\bullet$ & 0.1042 $\bullet$ & 0.17 & 1.2 $\times$ \\
		& BANE & 0.3165 $\bullet$ & 0.0217 $\bullet$ & 0.2887 $\bullet$ & 0.0369 $\bullet$ & 0.2467 $\bullet$ & 0.0692 $\bullet$ & 0.17 & 1.2 $\times$\\
		& DNE & 0.1163 $\bullet$ & 0.0023 $\bullet$ & 0.1160 $\bullet$ & 0.0041 $\bullet$ & 0.1157 $\bullet$ & 0.0091 $\bullet$ & 0.13 & 0.9 $\times$ \\
		& KD-coding & 0.1923 $\bullet$ & 0.0076 $\bullet$ & 0.1817 $\bullet$ & 0.0127 $\bullet$ & 0.1655 $\bullet$ & 0.0242 $\bullet$ & 0.13 & 0.9 $\times$ \\
		& NetHash & 0.2035  $\bullet$ & 0.0090  $\bullet$ & 0.1814  $\bullet$ & 0.0133  $\bullet$ & 0.1594  $\bullet$ & 0.0232  $\bullet$ & 4.28 & 30.5 $\times$\\
		& BinaryNE\_S & 0.2219 $\bullet$ & 0.0104 $\bullet$ & 0.2071 $\bullet$ & 0.0175 $\bullet$ & 0.1874 $\bullet$ & 0.0339 $\bullet$ & 0.13 & 0.9 $\times$ \\
		& BinaryNE\_C & \underline{0.5469} $\bullet$ & \underline{0.0486} $\bullet$ & \underline{0.4897} $\bullet$ & \underline{0.0805} $\bullet$ & 0.4010 $\bullet$ & 0.1434 $\bullet$ & 0.14 & 1.0 $\times$ \\
		& BinaryNE\_S+C & 0.4293 $\bullet$ & 0.0322 $\bullet$ & 0.3813 $\bullet$ & 0.0523 $\bullet$ & 0.3125 $\bullet$ & 0.0918 $\bullet$ & 0.14 & 1.0 $\times$ \\
		& BinaryNE & \textbf{0.5865} $\;\;$ & \textbf{0.0548} $\;\;$ & \textbf{0.5340} $\;\;$ & \textbf{0.0938} $\;\;$ & \textbf{0.4467} $\;\;$ & \textbf{0.1747} $\;\;$ & 0.14 & \\
		\hline
	\end{tabular}
	\label{Res_Flickr}
\end{table*}

\begin{table*}[t]
	\centering
	\scriptsize
	\tabcolsep 2pt
	\caption{Similarity search results on DBLP(Subgraph)}
	\renewcommand{\arraystretch}{1}
	\begin{tabular}{cccccccccc} 
		\hline
		Metric & Method & precision@100 & MAP@100 & precision@200 & MAP@200 & precision@500 & MAP@500 & Query time (ms) & Speedup\\\hline
		\multirow{9}{*}{Euclidean} 
		& DeepWalk & 0.7109 $\bullet$ & 0.0113 $\bullet$ & 0.6941 $\bullet$ & 0.0214 $\bullet$ & 0.6648 $\bullet$ & 0.0488 $\bullet$ & 11.07 & 31.6 $\times$ \\
		& LINE1 & 0.6852 $\bullet$ & 0.0106 $\bullet$ & 0.6421 $\bullet$ & 0.0190 $\bullet$ & 0.5338 $\bullet$ & 0.0350 $\bullet$ & 11.07 & 31.6 $\times$ \\
		& LINE2 & 0.6302 $\bullet$ & 0.0091 $\bullet$ & 0.5720 $\bullet$ & 0.0151 $\bullet$ & 0.4800 $\bullet$ & 0.0273 $\bullet$ & 11.07 & 31.6 $\times$ \\
		& SDNE & 0.6464 $\bullet$ & 0.0096 $\bullet$ & 0.6033 $\bullet$ & 0.0172 $\bullet$ & 0.5149 $\bullet$ & 0.0332 $\bullet$ & 11.07 & 31.6 $\times$ \\
		& TADW & 0.7156 $\bullet$ & 0.0107 $\bullet$ & 0.6951 $\bullet$ & 0.0199 $\bullet$ & 0.6581 $\bullet$ & 0.0441 $\bullet$ & 11.07 & 31.6 $\times$ \\
		& UPP-SNE & 0.7164 $\bullet$ & 0.0115 $\bullet$ & 0.7098 $\bullet$ & 0.0224 $\bullet$ & 0.6950 $\bullet$ & 0.0535 $\bullet$ & 11.07 & 31.6 $\times$ \\
		& MVC-DNE & 0.5575 $\bullet$ & 0.0067 $\bullet$ & 0.5285 $\bullet$ & 0.0118 $\bullet$ & 0.4908 $\bullet$ & 0.0245 $\bullet$ & 11.07 & 31.6 $\times$ \\
		& SINE & 0.7364 $\bullet$ & 0.0118 $\bullet$ & 0.7130 $\bullet$ & 0.0219 $\bullet$ & 0.6714 $\bullet$ & 0.0480 $\bullet$ & 11.07 & 31.6 $\times$ \\
		& Feature & 0.4555 $\bullet$ & 0.0043 $\bullet$ & 0.4408 $\bullet$ & 0.0077 $\bullet$ & 0.4289 $\bullet$ & 0.0172 $\bullet$ & 202.93 & 579.8 $\times$ \\
		\hline
		\multirow{17}{*}{Hamming} 
		& DeepWalk+Q & 0.7261 $\bullet$ & 0.0117 $\bullet$ & 0.7155 $\bullet$ & 0.0225 $\bullet$ & 0.6960 $\bullet$ & 0.0528 $\bullet$ & 0.35 & 1.0 $\times$ \\
		& LINE1+Q & 0.6881 $\bullet$ & 0.0106 $\bullet$ & 0.6528 $\bullet$ & 0.0193 $\bullet$ & 0.5555 $\bullet$ & 0.0366 $\bullet$ & 0.36 & 1.0 $\times$ \\
		& LINE2+Q & 0.6682 $\bullet$ & 0.0099 $\bullet$ & 0.6278 $\bullet$ & 0.0173 $\bullet$ & 0.5553 $\bullet$ & 0.0340 $\bullet$ & 0.38 & 1.1 $\times$ \\
		& SDNE+Q & 0.6922 $\bullet$ & 0.0103 $\bullet$ & 0.6740 $\bullet$ & 0.0195 $\bullet$ & 0.6256 $\bullet$ & 0.0425 $\bullet$ & 0.35 & 1.0 $\times$ \\
		& TADW+Q & 0.6964 $\bullet$ & 0.0111 $\bullet$ & 0.6751 $\bullet$ & 0.0206 $\bullet$ & 0.6329 $\bullet$ & 0.0444 $\bullet$ & 0.35 & 1.0 $\times$ \\
		& UPP-SNE+Q & 0.7291 $\bullet$ & \underline{0.0120} $\bullet$ & \underline{0.7256} $\bullet$ & \underline{0.0238} $\bullet$ & \textbf{0.7182} $\;\;$ & \textbf{0.0582} $\circ$ & 0.36 & 1.0 $\times$ \\
		& MVC-DNE+Q & 0.5234 $\bullet$ & 0.0057 $\bullet$ & 0.5005 $\bullet$ & 0.0101 $\bullet$ & 0.4710 $\bullet$ & 0.0217 $\bullet$ & 0.36 & 1.0 $\times$ \\
		& SINE+Q & \underline{0.7395} $\bullet$ & \underline{0.0120} $\bullet$ & 0.7207 $\bullet$ & 0.0227 $\bullet$ & 0.6867 $\bullet$ & 0.0510 $\bullet$ & 0.38 & 1.1 $\times$ \\
		& Feature+Q & 0.5489 $\bullet$ & 0.0067 $\bullet$ & 0.5304 $\bullet$ & 0.0122 $\bullet$ & 0.4983 $\bullet$ & 0.0263 $\bullet$ & 0.36 & 1.0 $\times$ \\
		& BANE & 0.7285 $\bullet$ & 0.0116 $\bullet$ & 0.7032 $\bullet$ & 0.0214 $\bullet$ & 0.6608 $\bullet$ & 0.0465 $\bullet$ & 0.38 & 1.1 $\times$\\
		& DNE & 0.6237 $\bullet$ & 0.0085 $\bullet$ & 0.5980 $\bullet$ & 0.0155 $\bullet$ & 0.5695 $\bullet$ & 0.0348 $\bullet$ & 0.36 & 1.0 $\times$ \\
		& KD-coding & 0.6437 $\bullet$ & 0.0085 $\bullet$ & 0.6683 $\bullet$ & 0.0182 $\bullet$ & 0.6586 $\bullet$ & 0.0453 $\bullet$ & 0.29 & 0.8 $\times$ \\
		& NetHash& 0.6606 $\bullet$ & 0.0097 $\bullet$ & 0.6242 $\bullet$ & 0.0171 $\bullet$ & 0.5750 $\bullet$ & 0.0357 $\bullet$ & 7.31 & 20.9 $\times$\\
		& BinaryNE\_S & 0.7195 $\bullet$ & 0.0115 $\bullet$ & 0.7069 $\bullet$ & 0.0222 $\bullet$ & 0.6830 $\bullet$ & 0.0518 $\bullet$ & 0.32 & 0.9 $\times$ \\
		& BinaryNE\_C & 0.5500 $\bullet$ & 0.0064 $\bullet$ & 0.5329 $\bullet$ & 0.0118 $\bullet$ & 0.5088 $\bullet$ & 0.0261 $\bullet$ & 0.35 & 1.0 $\times$ \\
		& BinaryNE\_S+C & 0.7353 $\bullet$ & 0.0119 $\bullet$ & 0.7203 $\bullet$ & 0.0228 $\bullet$ & 0.6934 $\bullet$ & 0.0520 $\bullet$ & 0.37 & 1.1 $\times$ \\
		& BinaryNE & \textbf{0.7528} $\;\;$ & \textbf{0.0125} $\;\;$ & \textbf{0.7403} $\;\;$ & \textbf{0.0242} $\;\;$ & \underline{0.7155} $\;\;$ & \underline{0.0563} $\;\;$ & 0.35 & \\
		\hline
	\end{tabular}
	\label{Res_DBLP(Subgraph)}
\end{table*}

From Tables~\ref{Res_Cora}-\ref{Res_DBLP(Subgraph)}, we can see that BinaryNE achieves significantly better precision and MAP than other discrete embedding methods \dk{on Cora, Citeseer, Flickr, and DBLP(Subgraph), except for MAP@100, MAP@200 on Cora, and MAP@500 on DBLP(Subgraph) that are slightly worse than the best performers. 
On BlogCatalog, among the discrete embedding methods, TADW+Q performs best while BinaryNE follows with comparable performance. From the above observations, we can see that our BinaryNE achieves the best overall similarity search performance on all datasets among the discrete embedding methods. What also deserves to be noticed is that BinaryNE still delivers competitive or even better performance compared with the continuous network embedding methods. Through using node binary representations to predict the existence of context nodes, nodes sharing similar structural context are embedded closely in the binary space. Similar binary representations are also learned for nodes sharing similar node attributes with the same mechanism, by using node binary representations to predict node attributes. The seamless integration between embedding binarization and embedding learning empowers BinaryNE to effectively capture both network structure and node content features, which makes the learned binary codes informative enough to measure node similarity accurately.}

On the other hand, BinaryNE remarkably improves search efficiency, providing more than 25 times faster search speed than continuous network embedding methods, and more than 20 times than NetHash, which learns non-binary discrete representations. Compared with the Euclidean distance in the continuous embedding space and the Hamming distance in the non-binary discrete embedding space, the Hamming distance measured by binary representations can be calculated far more efficiently with the bit-wise operations. 

Among the continuous network embedding baselines, the attributed network embedding (TADW, UPP-SNE or SINE) achieves the best search precisions. On the five networks, raw node content features consistently fail to achieve satisfactory precisions. By integrating network structure and node content in measuring node similarity, attributed network embedding is superior to structure preserving network embedding and raw node content features.

\dk{With the state-of-the-art quantization technique, the binarized continuous network embeddings are able to achieve satisfactory search precision in some cases. However, they still tend to be inferior when they are compared with our BinaryNE algorithm.} The results demonstrate that it is suboptimal to separately learn continuous network embeddings and quantize them into binary codes. In comparison, BinaryNE directly encodes network structure and node content features into binary node representations, achieving superior search precisions.


BANE achieves relatively good performance on BlogCatalog, Flick and DBLP(Subgraph), but yields unsatisfactory performance on Cora and Citeseer. This is because, BANE performs neighboring node attribute vector aggregation to integrate network structure and node attributes into a unified binary node representation. When the inconsistency between network structure and attributes occurs, \textit{i.e.}, linked nodes have discrepant attributes, the similarity measured by network structure and node attributes tends to heavily deteriorate each other.

\dk{DNE and KD-coding fail to achieve satisfactory performance in most cases. They learn binary node representations from only network structure and do not leverage the essential information on node attributes. Without the supervision of node labels, DNE cannot learn discriminative binary codes, preventing it from performing well for similarity search.}

NetHash constructs discrete node representations by randomly sampling the IDs of node features aggregated from local neighborhood. With both network structure and node features leveraged, NetHash achieves relatively good performance on Cora, Citeseer, and DBLP(Subgraph). As the discrete embeddings do not take binary values, bit-wise operations cannot be performed to calculate Hamming distance. As a result, its query speedup over continuous network embedding is limited.

By comparing BinaryNE with its three ablated versions, BinaryNE\_S and BinaryNE\_C with only network structure or only node content preserved, as well as BinaryNE\_S+C with only half dimensions encoded with network structure/node content, we can find that BinaryNE consistently outperforms the three counterparts in all cases. This verifies 
BinaryNE's advantage of integrating both network structure and node content to learn informative binary node representations over the counterparts using only network structure or node content, as well as the naive combination of their respective binary node representations. 
Apart from Cora and Flickr, where network structure and node content respectively dominate similarity search, BinaryNE\_S+C indeed outperforms BinaryNE\_S and BinaryNE\_C on Citeseer, BlogCatalog and DBLP(Subgraph), with more information leveraged. However, as BinaryNE\_S+C encodes network structure/node content into only half dimensions of binary node representations, it is inferior to BinaryNE that preserves network structure and node content at each dimension.

\begin{table*}[t]
	\begin{minipage}[b]{0.5\textwidth} 
		\centering
		\scriptsize
		\tabcolsep 4pt
		\caption{Node classification accuracy on DBLP(Subgraph)}
		\renewcommand{\arraystretch}{1.3}
		\begin{tabular}{cccccccccc} 
			\hline
			Training Ratio & 10\% & 30\% & 50\% & 70\% \\\hline
			DeepWalk & 0.7869 $\bullet$ & 0.7997 $\bullet$ & 0.8024 $\bullet$ & 0.8036 $\bullet$\\
			LINE1 & 0.7320 $\bullet$ & 0.7572 $\bullet$ & 0.7620 $\bullet$ & 0.7639 $\bullet$\\
			LINE2 & 0.6567 $\bullet$ & 0.6941 $\bullet$ & 0.7026 $\bullet$ & 0.7089 $\bullet$\\
			SDNE & 0.6391 $\bullet$ & 0.6676 $\bullet$ & 0.6782 $\bullet$ & 0.6838 $\bullet$\\
			TADW & 0.7903 $\bullet$ & 0.8125 $\bullet$ & 0.8186 $\bullet$ & 0.8217 $\;$\\
			UPP-SNE & 0.7881 $\bullet$ & 0.7965 $\bullet$ & 0.7977 $\bullet$ & 0.7993 $\bullet$\\
			MVC-DNE & 0.6984 $\bullet$ & 0.7478 $\bullet$ & 0.7589 $\bullet$ & 0.7650 $\bullet$\\
			SINE & \textbf{0.8054} $\circ$ & \textbf{0.8270} $\circ$ & \textbf{0.8319} $\circ$ & \textbf{0.8327} $\circ$\\
			BinaryNE & \underline{0.8005} $\;$ & \underline{0.8193} $\;$ & \underline{0.8216} $\;$ & \underline{0.8237} $\;$\\\hline
		\end{tabular}
		\label{Res_DBLP_classification}
	\end{minipage}
	\begin{minipage}[b]{0.45\textwidth} 
		\centering
		\scriptsize
		\tabcolsep 4pt
		\caption{Node clustering results on Cora}
		\renewcommand{\arraystretch}{1.3}
		\begin{tabular}{cccccccccc} 
			\hline
			Metric & Accuracy & Fvalue & NMI \\\hline
			DeepWalk & 0.6263 $\bullet$ & 0.6096 $\bullet$ & 0.4306 $\bullet$\\
			LINE1 & 0.3510 $\bullet$ & 0.3110 $\bullet$ & 0.1023 $\bullet$\\
			LINE2 & 0.4170 $\bullet$ & 0.3797 $\bullet$ & 0.1763 $\bullet$\\
			SDNE & 0.4061 $\bullet$ & 0.3804 $\bullet$ & 0.1939 $\bullet$\\
			TADW & 0.3705 $\bullet$ & 0.3556 $\bullet$ & 0.1487 $\bullet$\\
			UPP-SNE & 0.6337 $\bullet$ & 0.6209 $\bullet$ & 0.4412 $\bullet$\\
			MVC-DNE & 0.6095 $\bullet$ & 0.5885 $\bullet$ & 0.3510 $\bullet$\\
			SINE & \underline{0.6345} $\bullet$ & \underline{0.6239} $\bullet$ & \underline{0.4540} $\bullet$\\
			BinaryNE & \textbf{0.6729} $\;$ & \textbf{0.6656} $\;$ & \textbf{0.4810} $\;$\\
			\hline
		\end{tabular}
		\label{Res_Cora_Clustering}
	\end{minipage}
\end{table*}

\subsection{Experiments on Node Classification and Node Clustering}

We further evaluate the effectiveness of the binary codes learned by BinaryNE through node classification and node clustering. Although the binary codes learned by BinaryNE aim to serve efficient node similarity search, they can also speed up other tasks, like node classification, link prediction and node clustering. Here, with the guaranteed efficiency, we select node classification and node clustering to check whether BinaryNE is able to deliver satisfactory results for other downstream tasks, like the competitive continuous network embedding methods.  

We perform node classification experiments on DBLP(Subgraph). By taking the binary codes learned by BinaryNE and node representations learned by other continuous network embedding algorithms as features, we evaluate their classification effectiveness with a training-test split. We vary the training ratio from 10\% to 30\%, 50\%, and 70\%. Node classification accuracy values are reported in Table~\ref{Res_DBLP_classification}\footnote{In Tables 7-8, the best and the second best performers are highlighted by \textbf{bold} and \underline{underline}, respectively, and $\bullet$ ($\circ$) indicates that BinaryNE is significantly better (worse) than the compared baseline methods at $95\%$ significance level.\label{tables:7-8}}. As shown in Table~\ref{Res_DBLP_classification}, SINE performs the best while BinaryNE achieves the second best performance, which is comparable to the best performer. 

On Cora, we conduct node clustering experiments on the learned node representations with the \textit{k-means} algorithm. For the node representations learned by continuous network embedding algorithms, we use the Euclidean distance metric, while hamming distance is used for binary codes learned by BinaryNE, which is much more efficient than the Euclidean distance. Table~\ref{Res_Cora_Clustering}\footnotemark[\value{footnote}] reports the node clustering results on Accuracy, Fvalue, and NMI~\cite{Strehl03cluster}. From Table~\ref{Res_Cora_Clustering}, we can see that BinaryNE achieves the best clustering performance, significantly outperforming other continuous network embedding methods.

The excellent performance on node classification and node clustering proves that the binary codes learned by BinaryNE are informative enough to well represent network nodes. The BinaryNE binary codes can not only serve fast node similarity search with high search precision, but also underpin the fast operation of other network analytic tasks with competitive performance. 

\subsection{A Case Study on Relevant Paper Search}

In this subsection, we conduct a case study on relevant paper search on the DBLP network. We select the paper ``Learning Classifiers from Only Positive and Unlabeled Data" published on KDD-2008 as the query paper, which is a highly cited paper on the topic of ``Positive Unlabeled Learning". We retrieve the top-5 similar papers with the node representations learned by DeepWalk+Q, Feature+Q, TADW+Q, NetHash and BinaryNE, by calculating the Hamming distance between the query paper and candidate papers. Table~\ref{Res_paperSearch} reports the search results. As can be seen, DeepWalk+Q, Feature+Q, TADW+Q, and NetHash only retrieve one relevant paper, and no relevant papers are discovered by Feature+Q. By contrast, our BinaryNE algorithm achieves the best search results, with two relevant papers (1 and 5) discovered. It is worth noting that, as different algorithms use various information sources to measure node similarity, the top 5 relevant papers retrieved by different algorithms may have few intersections.

\begin{table}[t]
	\centering
	\scriptsize
	\tabcolsep 5pt
	\caption{Top-5 relevant paper search on DBLP}
	\renewcommand{\arraystretch}{1.15}
	\begin{tabular}{l} 
		\hline
		\textbf{Query}:  \textbf{Learning Classifiers from Only Positive and Unlabeled Data}\\\hline
		\textbf{DeepWalk+Q}:\\
		1. Finding Transport Proteins in a General Protein Database\\
		2. A Bayesian Network Framework for Reject Inference\\
		3. Making Generative Classifiers Robust to Selection Bias\\
		4. Building Text Classifiers Using \textbf{Positive and Unlabeled} Examples $\bm{\checkmark}$\\
		5. Audience Selection for On-line Brand Advertising: Privacy-friendly Social Network Targeting\\\hline
		\textbf{Feature+Q}:\\
		1. Learning Coordination Classifiers\\
		2. Learning from Little: Comparison of Classifiers Given Little Training\\
		3. Learning a Two-stage SVM/CRF Sequence Classifier\\
		4. Delegating Classifiers\\
		5. On the Chance Accuracies of Large Collections of Classifiers
		\\\hline
		\textbf{TADW+Q}:\\
		1. Efficient Learning of Naive Bayes Classifiers under Class-conditional Classification Noise\\
		2. Learning to Classify Texts Using \textbf{Positive and Unlabeled} Data $\bm{\checkmark}$\\
		3. Semi-Supervised Learning with Very Few Labeled Training Examples\\
		4. Calculation of the Learning Curve of Bayes Optimal Classification Algorithm for Learning a Perceptron With Noise\\
		5. How To Use What You Know\\\hline
		\textbf{NetHash}:\\
		1. Making Generative Classifiers Robust to Selection bias\\
		2. A Bayesian Network Framework for Reject Inference\\
		3. Building Text Classifiers Using \textbf{Positive and Unlabeled} Examples $\bm{\checkmark}$\\
		4. Finding Transport Proteins in a General Protein Database\\
		5. Active Learning in Partially Supervised Classification\\\hline
		\textbf{BinaryNE}:\\
		1. Learning to Classify Texts Using \textbf{Positive and Unlabeled} Data $\bm{\checkmark}$\\
		2. Learning the Common Structure of Data\\
		3. Enhancing Supervised Learning with Unlabeled Data\\
		4. Learning from Multiple Sources\\
		5. Text Classification from \textbf{Positive and Unlabeled} Documents $\bm{\checkmark}$\\\hline
	\end{tabular}
	\label{Res_paperSearch}
\end{table}

\subsection{Comparison of Memory Usage}

In Table~\ref{Res_storage}, we compare the memory used for accommodating the continuous node representations learned by DeepWalk, the non-binary discrete node representations learned by NetHash,  and the binary codes learned by BinaryNE. Compared with DeepWalk and NetHash, with the same dimension, the binary representations learned by BinaryNE significantly reduce the memory consumption by 64 and 32 times, respectively. For the DBLP(Full) network with more than 1 million nodes, the memory used for storing the continuous node representations is more than 1.5G, which is intractable for computing devices with low memory configuration to perform node similarity search. By contrast, the binary node representations learned by BinaryNE only consume 25M memory for the DBLP(Full) network, which is more practical for general devices. The low memory consumption makes BinaryNE more desirable for real-world applications.

\begin{table}[t]
	\centering
	\scriptsize
	\tabcolsep 4pt
	\caption{The memory usage of DeepWalk, NetHash and BinaryNE embeddings}
	\renewcommand{\arraystretch}{1.25}
	\begin{tabular}{cccccc} 
		\hline
		\multirow{2}{*}{Dataset} & \multicolumn{2}{c}{DeepWalk} & \multicolumn{2}{c}{NetHash} & BinaryNE \\
		& Memory &Reduction & Memory & Reduction & Memory \\\hline
		Cora & 2.64M  & 64$\times$ & 1.32M & 32$\times$ & 42.32K \\
		Citeseer & 3.23M & 64$\times$ & 1.62M & 32$\times$ & 51.75K\\
		BlogCatalog & 5.07M & 64$\times$ & 2.54M & 32$\times$ & 81.19K\\
		Flickr & 7.40M & 64$\times$  & 3.70M  & 32$\times$ & 118.36K \\
		DBLP(Subgraph) & 18.02 M & 64$\times$ & 9.01 M & 32$\times$ & 288.25 K\\
		DBLP(Full) & 1.56G & 64$\times$ & 797.09M & 32$\times$ & 24.91M\\\hline
	\end{tabular}
	\label{Res_storage}
\end{table}

\subsection{Experiments on Search Scalability}

\begin{figure}[t]
	\centering
	\subfigure[$|\mathcal{V}|$]{
		\centering
		\label{fig:scale:network_size} 
		\includegraphics[width=2.5in]{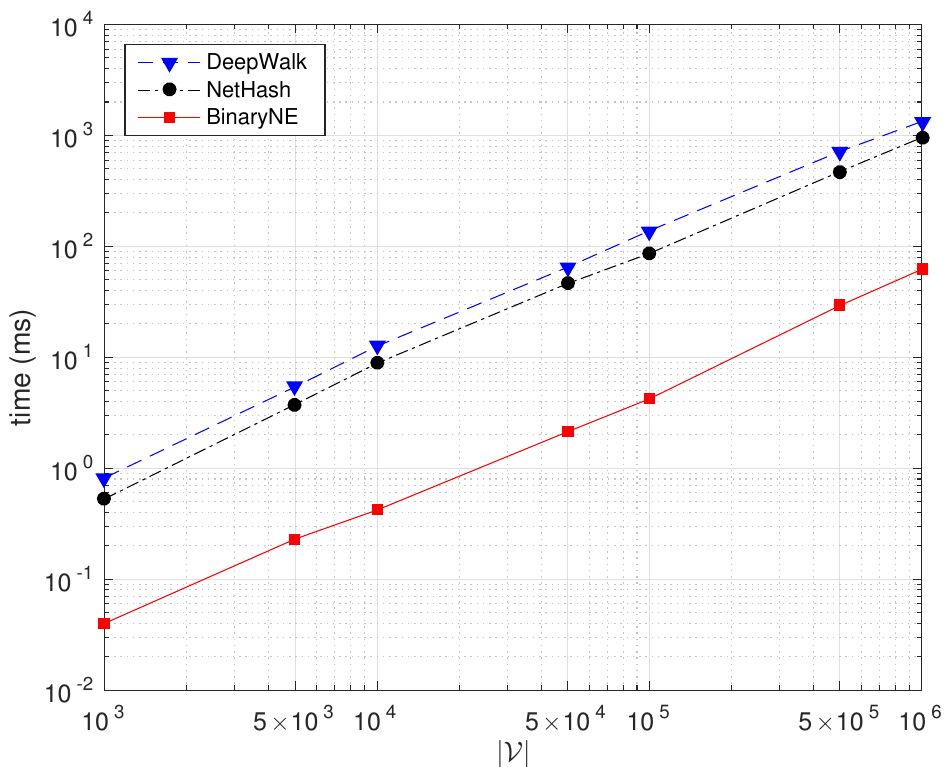}}
	\subfigure[$d$]{
		\label{fig:scale:dim} 
		\includegraphics[width=2.5in]{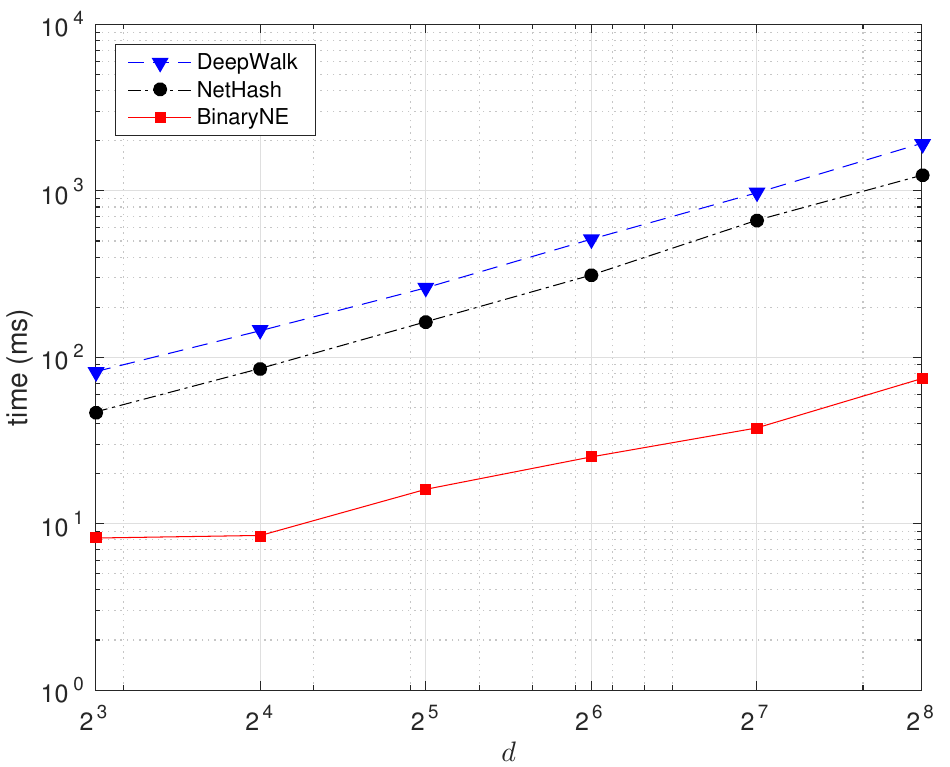}}
	\caption{Query time with varying $|\mathcal{V}|$ and $d$}
	\label{fig:scale} 
\end{figure}

We also conduct experiments on the large DBLP(Full) network to test the search scalability of different types of network embeddings with respect to network size $|\mathcal{V}|$ and embedding dimension $d$. We compare the binary embeddings generated by BinaryNE with those by DeepWalk and NetHash, which respectively take continuous numeric values and non-binary discrete values.

To study the search scalability on network size $|\mathcal{V}|$, we first learn 128-dimensional embeddings with DeepWalk, NetHash, and BinaryNE on the whole DBLP(Full) network, and then randomly sample a series of node subsets with increasing sizes. Among each node subset, we randomly select 1,000 nodes as query nodes and search similar nodes with the learned node representations. Fig.~\ref{fig:scale:network_size} shows query time (in milliseconds) with regard to different network sizes, where both query time (in milliseconds) and $|\mathcal{V}|$ are in logarithmic scales. As is shown, node similarity search with different embedding methods scales linearly with the increase of network size, whereas BinaryNE provides more than 10 times faster query speed than Deepwalk and NetHash. 

To study the search scalability in terms of embedding dimension $d$, we learn DeepWalk, NetHash and BinaryNE embeddings with varying dimensions (8, 16, 32, 64, 128 and 256). We randomly select 100 nodes as query nodes, and search similar nodes across the whole DBLP(Full) network. Fig.~\ref{fig:scale:dim} shows query time (in milliseconds) with varying embedding dimensions, with both axes in logarithmic scale. We can see that, in general, similarity search with three methods scales almost linearly with regards to embedding dimension, but BinaryNE is consistently more efficient than DeepWalk and Nethash (with more than 10 times search speedup in most cases).

\subsection{Comparison of Embedding Learning Time}

\begin{figure}[t]
	\centering
	\includegraphics[width=3.4in]{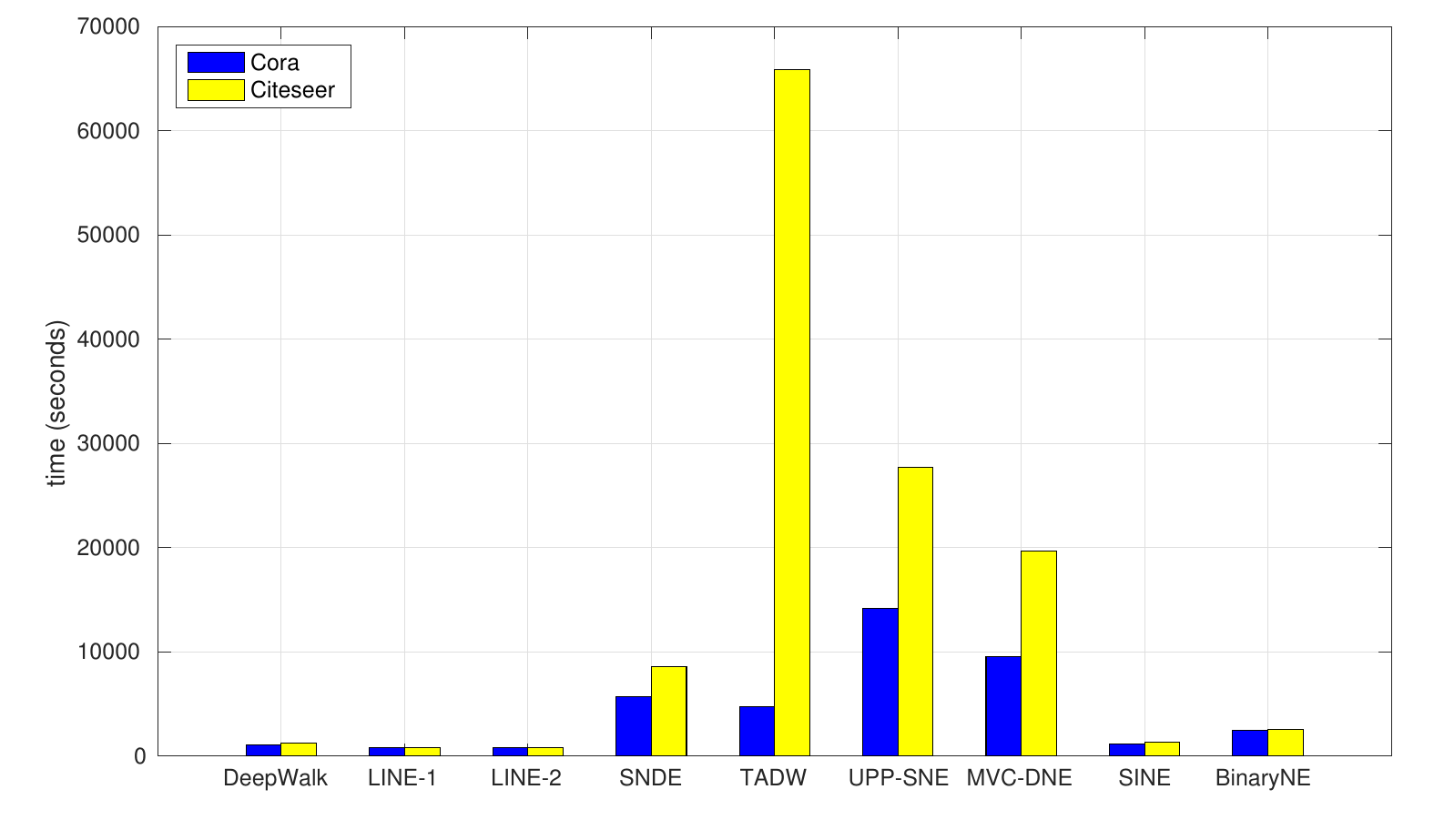}
	\caption{The time consumed by different network embedding methods for learning node representations}
	\label{fig:running_time} 
\end{figure}

We now select the Cora and Citeseer network to evaluate the efficiency of learning node representations with different network embedding methods. Fig.~\ref{fig:running_time} compares the CPU time (in seconds) consumed by different network embedding methods. As shown in the figure, BinaryNE is far more efficient in learning node representations than SDNE, TADW, UPP-SNE, and MVC-DNE, and its efficiency is comparable to that of DeepWalk, LINE1, LINE2, and SINE, which have been demonstrated to be efficient on large-scale networks. This proves the ability of BinaryNE to scale to large-scale networks for learning node representations, like DeepWalk, LINE and, SINE.

\begin{figure*}[t]
	\centering
	\subfigure[\#iteration]{
		\label{fig:para:subfig:iteration} 
		\includegraphics[width=1.75in]{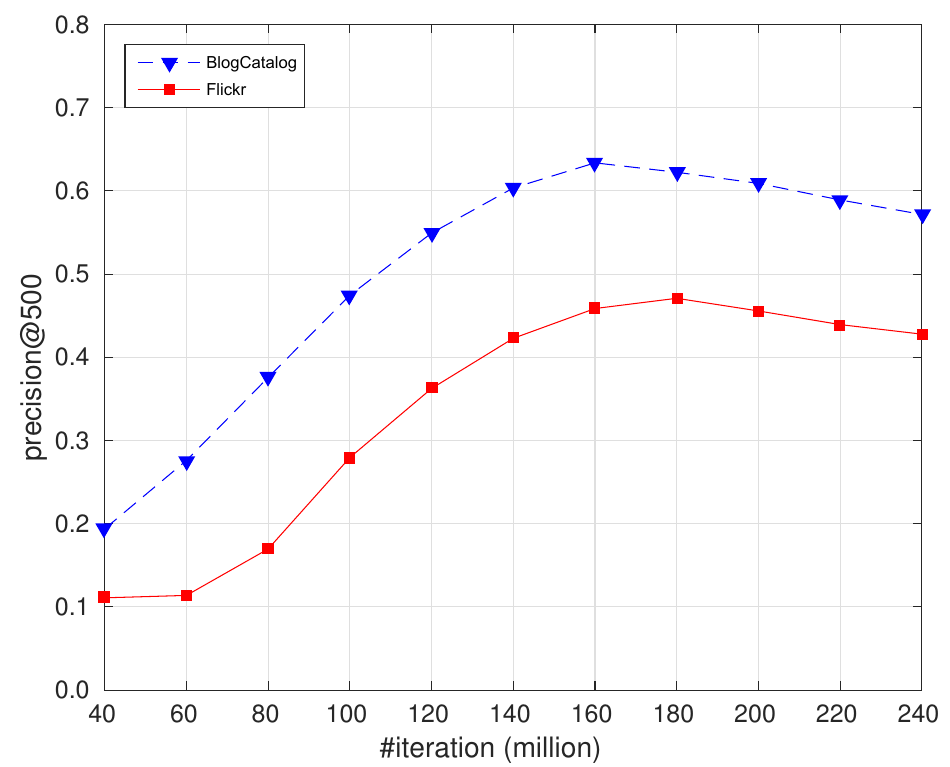}}
	\subfigure[$d$]{
		\label{fig:para:subfig:dim} 
		\includegraphics[width=1.75in]{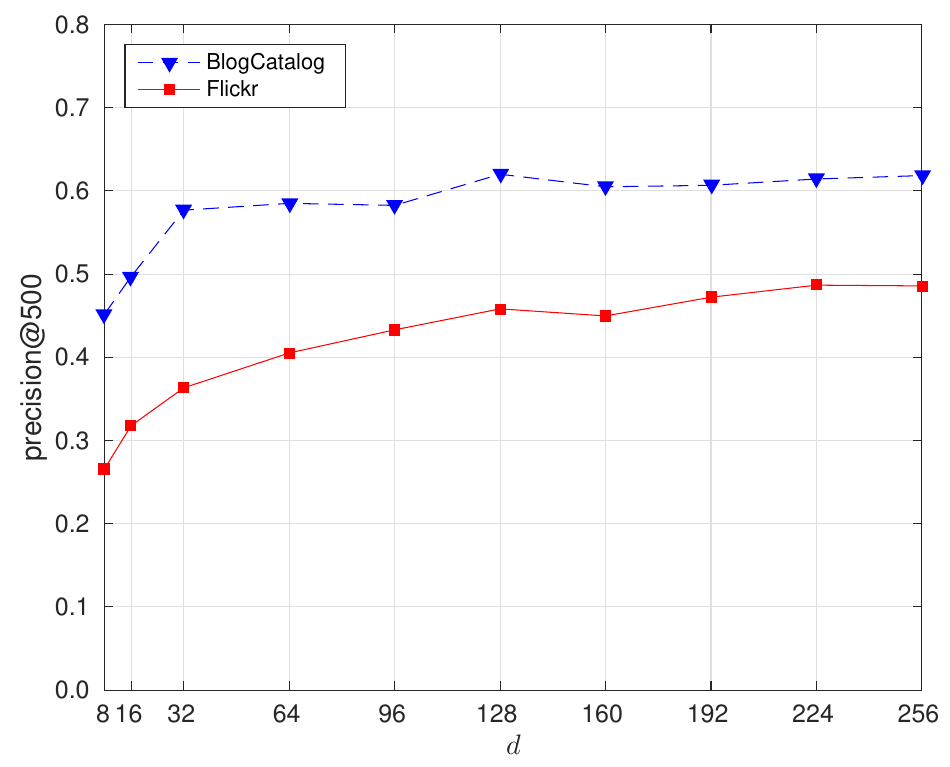}}
	\subfigure[$t$]{
		\label{fig:para:subfig:window} 
		\includegraphics[width=1.75in]{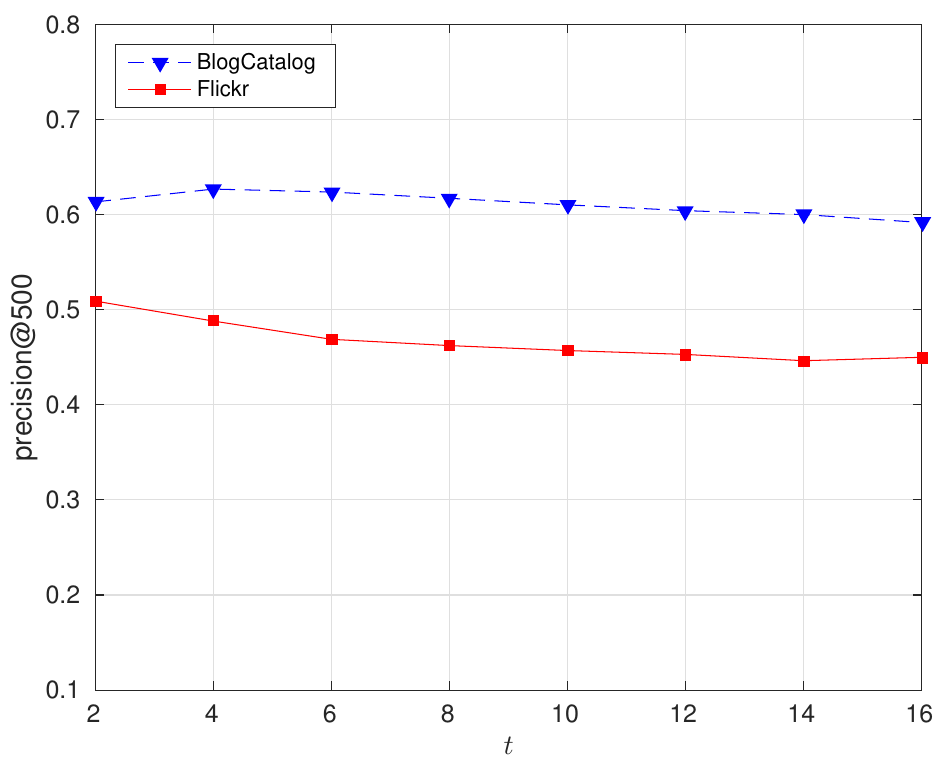}}
	\caption{The sensitivity of BinaryNE with parameters: the number of iterations, the dimension of learned embeddings $d$, and the window size $t$}
	\label{fig:para} 
\end{figure*}

\subsection{Experiments on Parameter Sensitivity}

Lastly, we perform a case study on BlogCatalog and Flickr to investigate the sensitivity of BinaryNE to three important parameters: the number of iterations, the dimension of learned embeddings $d$, and the window size $t$ used for collecting node context pairs. We take turns to fix any two parameters and study the effect of the remaining parameter on the search performance measured by precision@500. Fig.~\ref{fig:para} shows the performance of BinaryNE with respect to varying parameters. As the number of iterations increases, the performance of BinaryNE gradually increases and then declines slightly. This indicates that, in general, more iterations would be helpful for BinaryNE to find the local minimal solution, but excessive iterations tend to make the model parameters deviate from the local minima. \dk{However, to guarantee good performance, the number of iterations should scale linearly to the network scale. In practice, the number of iterations should be set to $500\sim1000$ times the sum of the number of edges and the number of node-attribute occurrence times.} When the embedding dimension $d$ increases, the performance of BinaryNE increases and stabilizes later. This shows that, embeddings with higher dimensions provide more information to measure node similarity. Interestingly, when the window size $t$ increases from 2 to 16, the search precision drops slightly. This is probably because a larger window size imports broader contextual structure, but may introduce more noise to measure node similarity.

\section{Conclusion}

Learning binary node representations is a desirable solution to similarity search over large-scale networks, due to its advantage in efficient bit-wise Hamming distance calculation and low memory usage. In this paper, we proposed a BinaryNE algorithm to embed network nodes into a binary space, with well preserved network structure and node content features. Through a three-layer neural network, BinaryNE learns binary node representations by modeling node structural context and node attribute relations. The \textit{sign} function is adopted as the activation function in the hidden layer to obtain binary node representations. To deal with the \textit{ill-posed gradient} problem caused by the non-smoothness of the \textit{sign} activation function, the state-of-the-art continuation technique~\cite{allgower2012numerical, cao2017hashnet} is employed. Model parameters are efficiently learned through an online stochastic gradient descent algorithm, which ensures the low time complexity and great scalability of BinaryNE. Extensive experiments on six real-world networks show that BinaryNE exhibits much lower memory usage and computational cost than continuous network embedding algorithms, but with comparable or even better search precision. \jy{The binary node representations learned by BinaryNE also achieve competitive results on other network analytic tasks such as node classification and node clustering.}

\section*{Acknowledgments}
The work is supported by a joint CRP research fund between the University of Sydney and Data61, CSIRO, the US National Science Foundation (NSF) through grants IIS-1763452, CNS-1828181, and the Australian Research Council (ARC) through grant LP160100630 and DP180100966. 

\bibliographystyle{ACM-Reference-Format}
\bibliography{BinaryNE}
\newpage
\section*{Appendix: Experimental Settings}

All reported experiments are conducted on a Mac laptop with an Inter Core i5, 2.3 GHz processor and 8 GB memory, with no GPU or other accelerators used. 

For all embedding learning methods used in our experiments, we set the dimension of embeddings $d=128$. For DeepWalk, UPP-SNE, SINE, and BinaryNE, we set the length of random walks $L=100$, the number of random walks starting from per node $\gamma=40$, and the window size $t=10$. 

For fair comparisons, we use the same strategy to train DeepWalk, UPP-SNE, SINE, and BinaryNE: we first collect node context pairs from the generated random walks, and update parameters with stochastic gradient descent by sampling node context pairs. We uniformly set the number of sampled negative context nodes $K$ to 5. For DeepWalk, LINE, UPP-SNE, SINE and BinaryNE, we set the maximum number of iterations to 100 million for Cora and Citeseer, 200 million for BlogCatalog, Flickr and DBLP(Subgraph). For DeepWalk and BinaryNE, we set the maximum number of iterations to 1 billion for DBLP. For DeepWalk, LINE, UPP-SNE, SINE, and BinaryNE, we gradually decrease the learning rate $\eta$ from 0.025 to $2.5\times10^{-6}$. We use the author provided LINE\footnote{https://github.com/tangjianpku/LINE}, UPP-SNE\footnote{https://github.com/daokunzhang/UPPSNE}, and SINE\footnote{https://github.com/daokunzhang/SINE} implementations. We implement DeepWalk based on the SINE implementation.

We use the SDNE implementation\footnote{https://github.com/suanrong/SDNE} provided by authors and implement MVC-DNE based on the SDNE implementation.  For SDNE, we find the best hyperparameter setting for $\alpha$, $\nu$, and $\beta$ via a grid search from $\{0.01, 0.1\}\times\{0.01, 0.1\}\times\{10, 50\}$ by evaluating 
on randomly selected 10\% nodes for each network. The number of neurons at each layer is set to 2708-512-128, 3312-512-128, 5,196-512-128, 7,575-512-128, and 18,448-512-128 for Cora, Citeseer, BlogCatalog, Flickr and DBLP(Subgraph), respectively. For MVC-DNE, on Cora, Citeseer, BlogCatalog, Flickr, and DBLP(Subgraph), the number of neurons at each layer in the structure view is set to 2708-512-64, 3312-512-64, 5,196-512-64, 7,575-512-64, and 18,448-512-64, respectively, and the number of neurons at each layer in the node content feature view is set to 1,433-512-64, 3,703-512-64, 8,189-512-64, 12,047-512-64, and 2,476-512-64, respectively. When setting the number of neurons at each layer for SDNE and MVC-DNE, we gradually decrease the number of neurons from the high-dimensional input layer to 128 neurons in the final node representation layer, so as to hierarchically extract deeper and more abstract latent features, as is operated in SDNE~\cite{wang2016structural} and MVC-DNE~\cite{yang2017properties}. For SDNE and MVC-DNE, 500 \dk{default} epochs are used for pre-training and parameter fine-tuning, respectively. Other parameters of SDNE and MVC-DNE are set according to~\cite{yang2017properties}.

As the content feature dimension of BlogCatalog, Flickr, and DBLP(Subgraph) is too large for TADW, before running TADW on them, we reduce the dimension of their node content features to 200 with SVD. Default settings are used to train NetHash and BANE. For BinaryNE, we gradually increase the parameter $\beta$ from 0.01 to 1. We use the author provided TADW\footnote{https://github.com/benedekrozemberczki/TADW}, BANE\footnote{https://github.com/shiruipan/BANE} and NetHash\footnote{https://github.com/williamweiwu/williamweiwu.github.io/tree/master/Graph\_Network\%20Embedding/NetHash} implementations. We implement the KD-coding for network embedding based on the original KD-coding implementation for word embedding\footnote{https://github.com/chentingpc/kdcode-lm}. We implement the DNE algorithm according to the DNE paper~\cite{xiaobo2018discrete}.

\end{document}